\documentclass[draft]{agujournal2019}
\usepackage{url} 
\usepackage{lineno}
\usepackage{soul}
\journalname{JGR: Space Physics}

\begin{document}

%
%


\title{Electrostatic Waves and Electron Heating Observed over Lunar Crustal Magnetic Anomalies}

%
%




\authors{F. Chu\affil{1,2}, J. S. Halekas\affil{1}, Xin Cao\affil{1}, J. P. McFadden\affil{3}, J. W. Bonnell\affil{3}, and K.-H. Glassmeier\affil{4}}


\affiliation{1}{Department of Physics and Astronomy, University of Iowa, Iowa City, IA, USA}
\affiliation{2}{Physics Division, Los Alamos National Laboratory, Los Alamos, NM, USA}
\affiliation{3}{Space Sciences Laboratory, University of California, Berkeley, CA, USA}
\affiliation{4}{Institut für Geophysik und Extraterrestrische Physik, Technische Universität Braunschweig, Braunschweig, Germany}




\correspondingauthor{F. Chu}{fchu@lanl.gov}




\begin{keypoints}
\item Two types of electrostatic instabilities are observed over the lunar crustal magnetic anomalies during ARTEMIS flyby
\item Electron two-stream instability and electron cyclotron drift instability may play an important role in driving the electrostatic waves
\item Electron cyclotron drift instability, along with modified two‐stream instability, may cause isotropic electron heating
\end{keypoints}

%
%


\begin{abstract}
Above lunar crustal magnetic anomalies, large fractions of solar wind electrons and ions can be scattered and stream back towards the solar wind flow, leading to a number of interesting effects such as electrostatic instabilities and waves. These electrostatic structures can also interact with the background plasma, resulting in electron heating and scattering. We study the electrostatic waves and electron heating observed over the lunar magnetic anomalies by analyzing data from the Acceleration, Reconnection, Turbulence, and Electrodynamics of Moon's Interaction with the Sun (ARTEMIS) spacecraft. Based on the analysis of two lunar flybys in 2011 and 2013, we find that the electron two-stream instability (ETSI) and electron cyclotron drift instability (ECDI) may play an important role in driving the electrostatic waves. We also find that ECDI, along with the modified two-stream instability (MTSI), may provide the mechanisms responsible for substantial isotropic electron heating over the lunar magnetic anomalies.
\end{abstract}

\section*{Plain Language Summary}
Without a global magnetic field or a thick atmosphere, the solar wind directly impacts the surface of the Moon. However, over regions where the lunar crust is strongly magnetized, the charged particles in the solar wind can be reflected and travel back towards the incoming solar wind, generating interesting features like electrostatic waves. These waves can also in turn affect the solar wind by increasing the temperature of its charged particles. To understand the mechanisms causing the waves and heating, we analyze data from the Acceleration, Reconnection, Turbulence, and Electrodynamics of Moon's Interaction with the Sun spacecraft. Our results indicate that the lunar environment becomes unstable because of the reflected charged particles, thereby creating free energies that lead to the waves and heating.

%
%

%


%
%
%
%

\section{Introduction}
\label{sec:intro}

In the absence of a global magnetic field and a thick atmosphere, unlike the case of the Earth, the surface of the Moon directly interacts with the incident solar wind plasma. Traditionally, the Moon has been thought to act as a simple barrier to the solar wind flow, causing the absorption of plasma at the upstream surface and formation of a plasma wake in the downstream. However, recent observations from Chandrayaan-1, Kaguya, and Chang’E-1 reveal that Moon-solar wind interaction is in fact much more complicated and dynamic, capable of creating a variety of interesting effects around the Moon. For example, the surface of the Moon, immersed in the solar wind plasma, charges to an electrostatic potential in order to balance the total incident currents \cite{whipple_potentials_1981, halekas_evidence_2002, halekas_new_2011}. Moreover, solar wind sputtering from the lunar surface and ionization of the tenuous neutral exosphere can produce heavier lunar pickup ions, which can then be accelerated downstream from the Moon by the motional electric field \cite{yokota_first_2009, halekas_lunar_2012-1, cao_acceleration_2020, cao_plasma_2020}. Some other examples of lunar interaction include backscattering of solar wind ions and photoelectron emission from the lunar surface \cite{reasoner_characteristics_1972, goldstein_observations_1974, lue_chandrayaan-1_2014, bhardwaj_new_2015, harada_photoemission_2017}.


One of the most interesting and unique Moon-solar wind interactions happens over the lunar crustal magnetic anomalies. Previous studies have shown that the lunar crustal magnetic fields can perturb the solar wind flow, causing plasma compressions at the lunar limb \cite{russell_source_1975}. More recent measurements from Kaguya suggest that ‘‘mini-magnetospheres’’ can form over strong magnetic anomaly regions, partially shielding the lunar surface from the solar wind \cite{saito_-flight_2010, wieser_first_2010, vorburger_energetic_2012}. In addition, local crustal magnetic fields are found capable of deflecting solar wind ions and electrons from the lunar surface \cite{lue_strong_2011, saito_simultaneous_2012, halekas_solar_2012, halekas_lunar_2012}. Using Chandrayaan-1 data, \citeA{lue_strong_2011} reported that on average $10 \%$ of the incident solar wind ions reflect over large-scale magnetic anomalies. The reflection efficiency can reach up to $50 \%$ for ions and as much as $100 \%$ for electrons above regions of strongest crustal fields \cite{halekas_mapping_2001}. The reflected ions and electrons, along with backscattered particles and photoelectrons, can then form counter-streaming ``beams'' towards the incoming solar wind flow, resulting in a number of fundamental plasma processes such as electrostatic instabilities and waves. These electrostatic structures can also in turn have an impact on the lunar plasma environment, leading to substantial electron heating and scattering. It should be noted that the reflected particles are not perfectly collimated beams; the ``beam'' model is just a first approximation that is useful for comparing to theory.


A variety of plasma instabilities and waves of different origin have been previously observed above the lunar crustal magnetic anomalies \cite{nakagawa_ulfelf_2016, harada_upstream_2016}. \citeA{tsugawa_statistical_2011} reported that monochromatic, left‐hand polarized (in the spacecraft frame) whistler waves with frequencies close to 1 Hz were detected by Kaguya near the Moon. A further statistical analysis suggested that the waves were generated by the solar wind interaction with lunar magnetic anomalies. In addition, broadband electrostatic mode, resulting from counter-streaming electron beams, is another type of waves commonly observed in the lunar upstream plasma  \cite{harada_extended_2014}.


In this paper, we investigate two types of electrostatic instabilities observed over the lunar crustal magnetic anomalies by Acceleration, Reconnection, Turbulence, and Electrodynamics of Moon's Interaction with the Sun (ARTEMIS) spacecraft. We report for the first time on a class of electrostatic waves propagating perpendicular to the ambient magnetic field, possibly driven by electron cyclotron drift instabilities. This type of electrostatic waves is analogous to those observed in the foot region of perpendicular shocks. In the end, we also discuss the mechanisms of electron heating observed along with the electrostatic waves over the magnetic anomalies.


\section{Instrumentation and Observations}
\label{sec:observations}

NASA's ARTEMIS spacecraft, consisting of two satellites (P1 and P2) originally from the THEMIS (Time History of Events and Macroscale Interactions During Substorms) mission, occupies stable 26-h period elliptical near-equatorial orbits around the Moon \cite{angelopoulos_artemis_2011}. To investigate the plasma environment above the dayside lunar surface, we utilize measurements from four of the instruments: the Electrostatic Analyzer \cite<ESA;>{mcfadden_themis_2008}, Electric Field Instrument \cite<EFI;>{bonnell_electric_2008}, Search Coil Magnetometer \cite<SCM;>{roux_search_2008}, and Fluxgate Magnetometer \cite<FGM;>{auster_themis_2008}. The ESA measures electron energies in the range of 2 eV to 32 keV and ion energies from 1.6 eV to 25 keV \cite{mcfadden_themis_2008}. The EFI is capable of measuring the three components of the ambient electric fields from $\sim 10$ Hz up to 8 kHz \cite{bonnell_electric_2008}.

%

\begin{figure*}
\begin{center}
\includegraphics[width=3.75in]{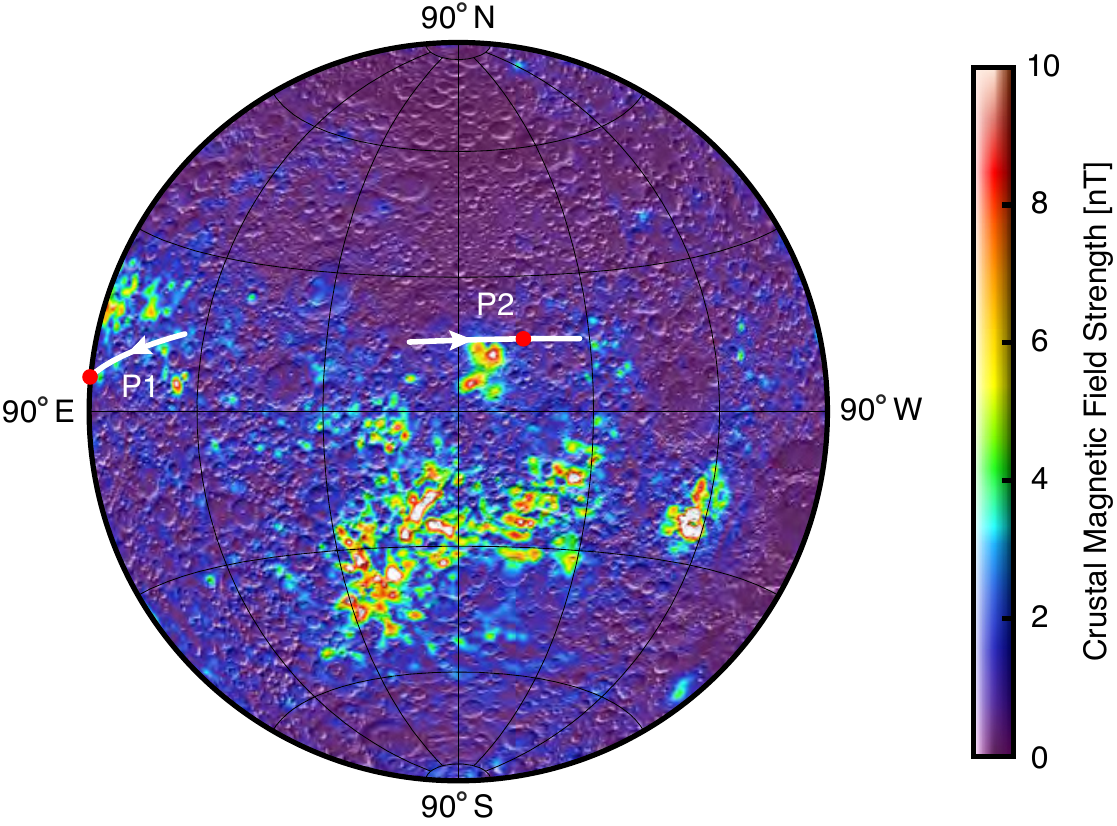}
\caption{Trajectories of ARTEMIS flybys over magnetic anomalies as shown in Figures~\ref{fig:case14} and \ref{fig:case5}, where the red dot indicates the periapsis. The crustal magnetic field strength map at 30 km altitude is based on data from the Tsunakawa crustal magnetic field model \cite{tsunakawa_surface_2015, wieczorek_strength_2018}.}
\label{fig:trajectories}
\end{center}
\end{figure*}
\begin{figure*}
\begin{center}
\includegraphics[width=6.69in]{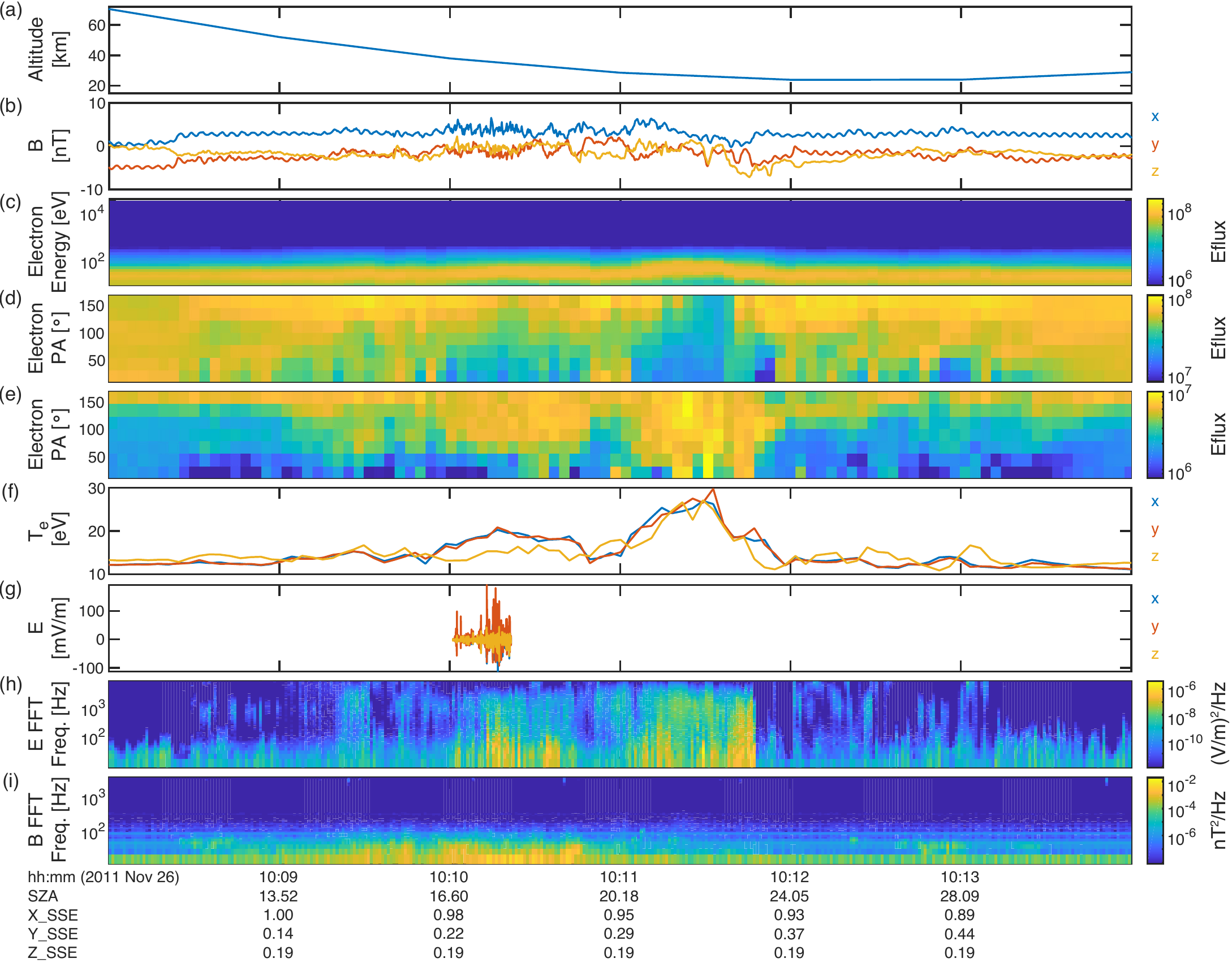}
\caption{Data from an ARTEMIS P2 lunar flyby over magnetic anomalies on 26 November 2011. (a) Altitude of the probe as a function of time. P2 reaches a periselene at an altitude of 23.8 km at 10:12 UT. (b) Ambient magnetic field in SSE coordinates. (c) Electron energy spectrogram. The differential energy flux has unites of eV/(eV $\textup{cm}^2$ sr s). (d)--(e) Electron pitch angle spectra for energies at 20 eV and 200 eV, respectively. (f) Electron temperatures parallel ($Z$ axis) and perpendicular ($X$ and $Y$ axis) to the magnetic field. (g) Wave burst data in magnetic field-aligned coordinates, $Z$ axis being parallel to the magnetic field. (h)--(i) FFT wave spectra of electric and magnetic field, respectively. Text labels indicate time of day in UT, solar zenith angle (SZA), and spacecraft (X, Y, Z) SSE coordinates in units of lunar radii ($R_L$).}
\label{fig:case14}
\end{center}
\end{figure*}

We select two cases best suited for studying the electrostatic waves and electron heating over the magnetic anomalies, where the observations can be made at altitudes below 50 km above the lunar surface. The trajectories of the ARTEMIS flybys, along with the crustal magnetic field strength map, are shown in Figure~\ref{fig:trajectories}. On 26 November 2011, ARTEMIS P2 flew by the Moon at average GSE coordinates of $[53,16,0]$ earth radii ($R_E$), located in the solar wind well upstream of the Earth’s bow shock. The data of the flyby obtained from the above four instruments are shown in Figure~\ref{fig:case14}. The probe is found to briefly fly over the magnetic anomaly region between 10:10 UT and 10:12 UT, indicated by both the trajectory in Figure~\ref{fig:trajectories} and an enhancement of the fluctuations in the ambient magnetic field in Figure~\ref{fig:case14}b. When the altitude of P2 descends below 50 km, two counter-streaming electron beams along the ambient magnetic field can be seen intermittently from the electron pitch angle spectra in Figures~\ref{fig:case14}d (energy around 20 eV) and \ref{fig:case14}e (energy around 200 eV). Since the magnetic field is $+B_x$ dominated in SSE coordinates, the $X$ axis being the direction pointing from the Moon toward the Sun, the beam with pitch angles centered around $180^\circ$ can therefore be identified as incoming solar wind electrons. The other beam centered at $\sim 0^\circ$ pitch angles results from the primary electrons reflected from the magnetic anomalies, as well as photoelectrons emitted from the dayside lunar surface \cite{whipple_potentials_1981, halekas_lunar_2012}.

During the time period of the flyby, we observe high frequency electrostatic fluctuations ranging from $\sim 0.1$ to 8 kHz, as shown in the electric field FFT spectrum in Figure~\ref{fig:case14}h. The fluctuations become more intense between 10:10 UT and 10:12 UT. Figure~\ref{fig:case14}g shows the high-resolution wave burst data (which has been rotated to magnetic field-aligned coordinates, with $E_z$ parallel to the field line), revealing that the electrostatic fluctuations are mostly perpendicular to the magnetic field between 10:10 UT and 10:11 UT. We also find broadband magnetic fluctuations extending from tens of Hz down to near-DC levels in magnetic field FFT spectrum (Figure~\ref{fig:case14}i). These waves are most likely to be whistler mode, as there are really no other electromagnetic modes that can propagate in this frequency range. Figure~\ref{fig:case14}f shows the electron temperatures parallel ($T_{e,z}$) and perpendicular ($T_{e,x}$ and $T_{e,y}$) to the magnetic field, where perpendicular electron heating is observed between 10:10 UT and 10:11 UT. In addition, strong isotropic heating is seen between 10:11 UT and 10:12 UT, accompanied by the intense electrostatic fluctuations. The electron heating can also be observed in the electron energy spectrogram in Figure~\ref{fig:case14}c.



\begin{figure*}
\begin{center}
\includegraphics[width=6.69in]{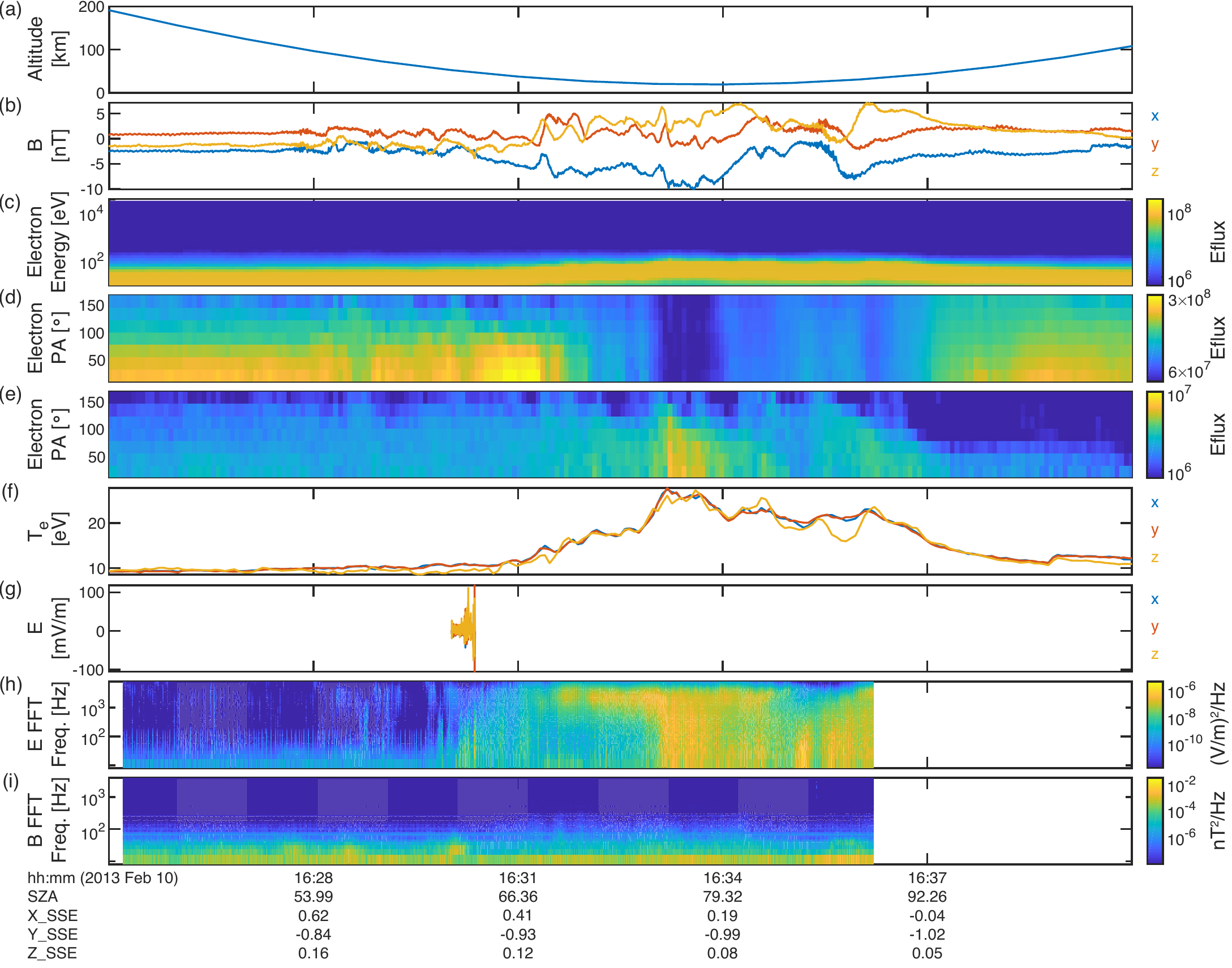}
\caption{Data from an ARTEMIS P1 lunar flyby over magnetic anomalies on 10 February 2013. P1 reaches the lowest altitude of 19.5 km above the lunar surface at 16:34 UT. All panels same as Figure~\ref{fig:case14}.}
\label{fig:case5}
\end{center}
\end{figure*}

Figure~\ref{fig:case5} shows an overview of another flyby (ARTEMIS P1) that occurred on 10 February 2013, when P1 was at average GSE coordinates of $[57,5,5]R_E$. The signatures we see are very similar to the previous event. Electrostatic fluctuations are observed in the electric field FFT spectrum (Figure~\ref{fig:case5}h) between 16:28 UT and 16:36 UT, although the field aligned wave burst data (Figure~\ref{fig:case5}g) indicate that the electrostatic fluctuations are mainly parallel to the magnetic field this time. Strong isotropic heating is also observed in the electron temperature profile (Figure~\ref{fig:case5}f) between 16:31 UT and 16:37 UT, coinciding with the intense electrostatic fluctuations. In addition, a recent analysis by \citeA{halekas_evidence_2014} pointed out that this flyby event has many of the aspects of a collisionless shock, despite the small scale size. 


\section{Origin of the Electrostatic Waves}
\label{sec:origin}

We now consider the generation mechanisms for the electrostatic fluctuations, which we suspect may result from a combination of different plasma processes in the complex lunar environment over the magnetic anomalies. The Moon acts as a barrier to the incoming solar wind flow. Due to the influence of crustal magnetic fields, large fractions of the solar wind electron and ion populations reflect above the lunar surface and stream towards the solar wind flow, resulting in varieties of plasma instabilities that could produce the electrostatic fluctuations shown in Section~\ref{sec:observations}. Two possible drivers for the waves in Figures~\ref{fig:case14} and \ref{fig:case5} are proposed: electron two-stream instability (ETSI) that could cause electrostatic fluctuations parallel to the ambient magnetic field, and electron cyclotron drift instability (ECDI), which can generate the electrostatic waves in the perpendicular direction.


\subsection{Electron Two-Stream Instability}
\label{subsec:streaming}


Electron two-stream instability driven by counter-streaming electron beams is one of the most commonly found electrostatic instabilities in space plasmas. For example, ETSI has been reported in the solar wind \cite{malaspina_electrostatic_2013}, Earth's magnetotail \cite{matsumoto_electrostatic_1994}, and at the bow shock \cite{bale_bipolar_1998}. The nonlinear evolution of ETSI often leads to the formation of time domain structures \cite{mozer_time_2015}, such as electrostatic solitary waves \cite{jao_fluid_2014, graham_electrostatic_2016}, electron phase-space holes \cite{franz_properties_2005, hutchinson_electron_2017, holmes_electron_2018, steinvall_multispacecraft_2019}, and double layers \cite{andersson_characteristics_2002, ergun_double_2003}.

\begin{figure*}
\begin{center}
\includegraphics[width=6.69in]{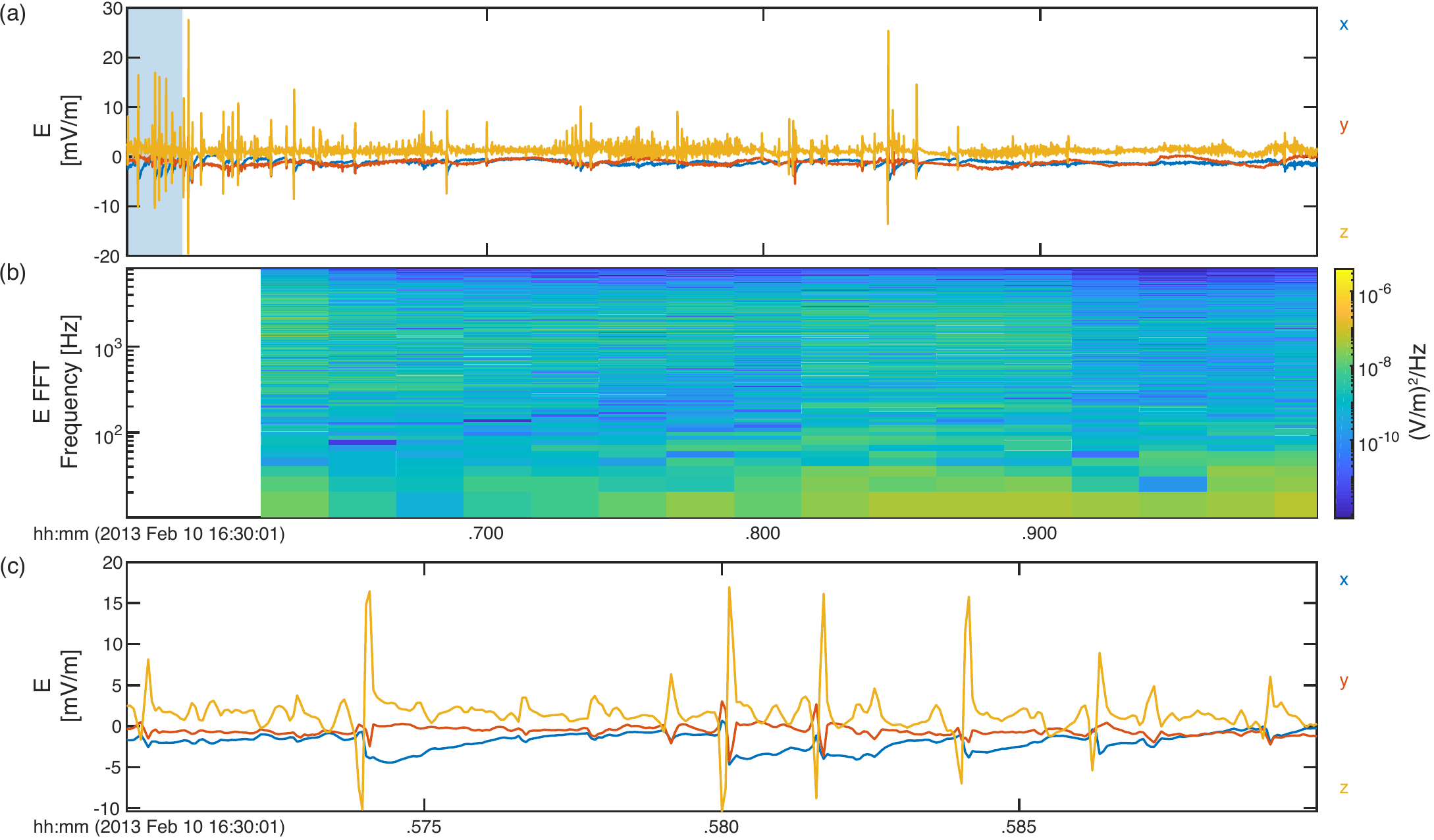}
\caption{(a) An example of the time domain structures observed during the ARTEMIS P1 lunar flyby on 10 February 2013. (b) Corresponding electric field FFT spectrum showing the broadband electrostatic fluctuations. (c) Zoom-in on the time scale over the blue-colored region in (a) to demonstrate electron phase-space holes.}
\label{fig:wave}
\end{center}
\end{figure*}

We show an example of the time domain structures observed during the ARTEMIS P1 lunar flyby in Figure~\ref{fig:wave}a and the corresponding electric field FFT spectrum in Figure~\ref{fig:wave}b. An upward electron beam can be observed at the same time in the electron pitch angle spectrum in Figure~\ref{fig:case5}d. Since the incident and counter-streaming electron beams are mostly adiabatic, the intense electric field spikes that result from the ETSI have significant field-aligned components $E_z$. If we zoom in on the time scale, a series of isolated bipolar electric field structures, known as electron phase-space holes, can be seen in Figure~\ref{fig:wave}c. Similar bipolar structures have also been previously observed near the Moon by Kaguya \cite{hashimoto_electrostatic_2010}. Electron phase-space holes are often responsible for heating and scattering background electrons through wave-particle interaction in many space plasmas \cite{mozer_magnetospheric_2016, vasko_diffusive_2017}.

\subsection{Electron Cyclotron Drift Instability}
\label{subsec:cyclotron}

Most of the electrostatic instabilities driven by either field-aligned beams or currents can only generate electrostatic fluctuations along the magnetic field. However, electron cyclotron drift instability, which arises from a relative drift between ions and electrons across the magnetic field, can lead to electrostatic waves in the electron cyclotron frequency in the perpendicular direction \cite{forslund_electron_1970}.



ECDI results from reactive coupling between electron cyclotron Bernstein modes and ion beam modes \cite{muschietti_microturbulence_2013}. The linear dispersion relation of the ECDI can be found in \citeA{janhunen_evolution_2018}. This type of instability has been observed in many laboratory plasmas \cite{ripin_electron_1973, stenzel_growth_1973} and occasionally in space \cite{wu_cyclotron_1972, wilson_large-amplitude_2010}. ECDI plays an important role in particle acceleration and heating in the foot of supercritical quasi-perpendicular shocks, where a fraction of incoming ions are reflected at the steep front \cite{matsukiyo_microinstabilities_2006}. Similar conditions favoring the occurrence of ECDI can be found in the lunar plasma environment tens of kilometers above the magnetic anomalies. The electrons in these regions are strongly magnetized; however, the ions are considered to be unmagnetized due to their very large gyroradii in comparison to the scale of the lunar crustal magnetic fields. The ion beam reflected from the magnetic anomalies therefore can stream in any direction, in particular, across the magnetic field, triggering the ECDI. 

\begin{figure*}
\begin{center}
\includegraphics[width=6.69in]{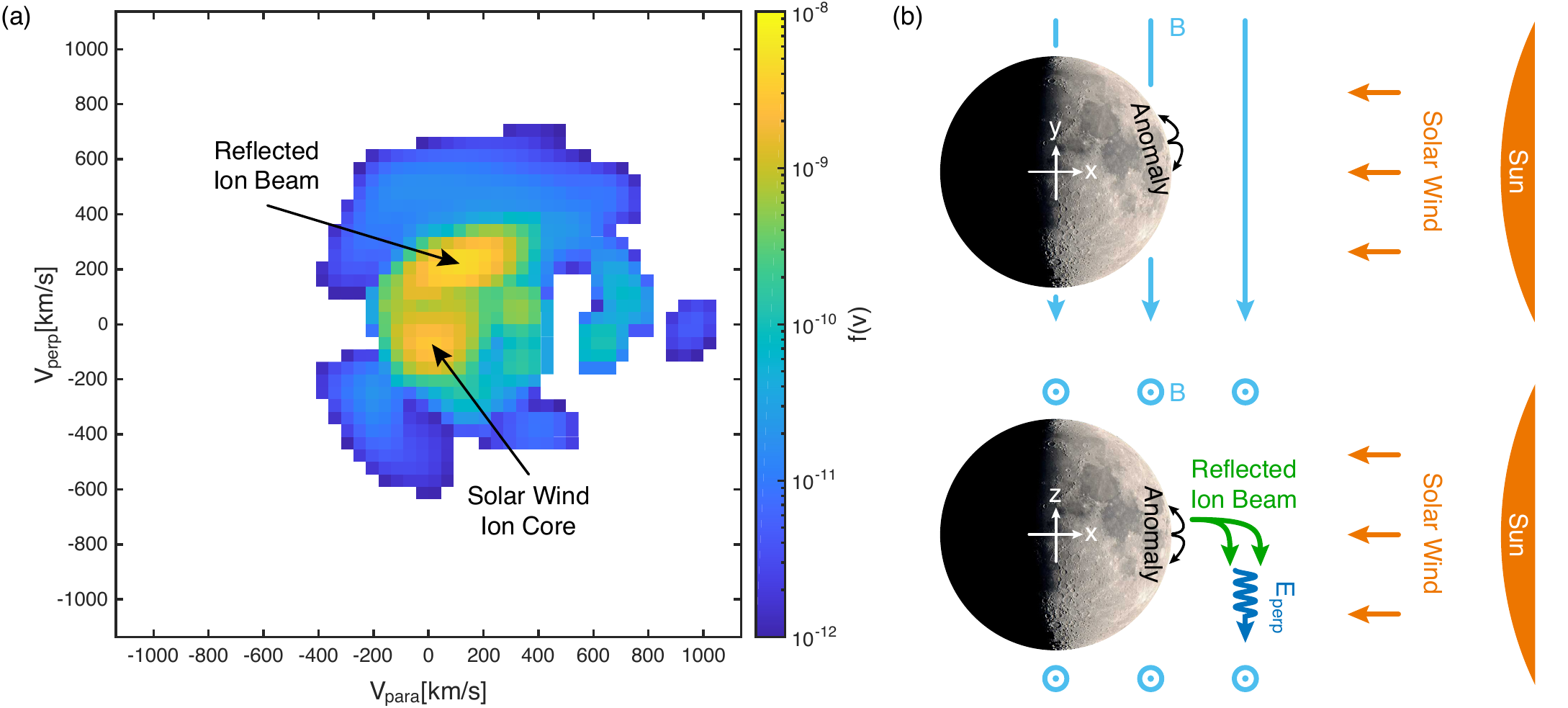}
\caption{(a) An example of a reflected ion beam traversing the solar wind plasma perpendicular to the background magnetic field. This sample ion velocity distribution cut, in plasma frame, was obtained at 10:11:30 UT during the P2 lunar flyby event as shown in Figure~\ref{fig:case14}. (b) Schematic illustration of the magnetic field ($-Y$ direction) and incoming solar wind ($-X$ direction) geometry during the flyby in SSE coordinates.}
\label{fig:ecdi}
\end{center}
\end{figure*}

ECDI provides a mechanism capable of driving the perpendicular electrostatic fluctuations during the P2 flyby as shown in Figure~\ref{fig:case14}g. Reflected ion beams traversing the solar wind plasma perpendicular to the background magnetic field are observed from ion velocity distributions, a good example of which is presented in Figure~\ref{fig:ecdi}a. Two ion beams can be correspondingly identified: a dense solar wind ion core close to the origin and the reflected ion beam streaming at $\sim 200$ km/s across the magnetic field. Figure~\ref{fig:ecdi}b illustrates the geometry of the magnetic field ($-Y$ direction) and incoming solar wind ($-X$ direction) during the flyby in SSE coordinates. Once the solar wind ions are reflected from the magnetic anomaly, they are accelerated by the motional electric field and stream towards the $-Z$ direction. The perpendicular configuration of the magnetic field and reflected ion beam therefore favors the occurrence of ECDI, resulting in the time-domain structures with perpendicular electric field in Figure~\ref{fig:case14}g. Further Fourier analysis indicates that these perpendicular electrostatic waves correspond to the strong wave power below $\sim 100$ Hz in Figure~\ref{fig:case14}h, within the frequency range of the electron cyclotron motion.

We now show that it is physically feasible for ECDI to excite the perpendicular electrostatic fluctuations under the observed conditions by comparing the wave growth time to the electron transit time across the altitude range ($\sim$ 50 km). The growth rate of ECDI can be estimated as \cite{wong_electrostatic_1970}
\begin{equation}
\label{eq:ECDI}
\gamma_n \approx \Omega _\textup{\scriptsize{ce}}\left ( \frac{1}{8 \pi} \frac{m_\textup{\scriptsize{e}}}{m_\textup{\scriptsize{i}}} \right )^{1/4}\frac{n^{1/2}}{\left ( 1+k^2 \lambda _\textup{\scriptsize{De}}^2 \right )^{3/4}},
\end{equation}
where $\Omega _\textup{\scriptsize{ce}}$ is the electron cyclotron frequency, $m_\textup{\scriptsize{e}}$ is the electron mass, $m_\textup{\scriptsize{i}}$ is the ion mass, $k \approx n\Omega _\textup{\scriptsize{ce}}/V_\textup{\scriptsize{b}}$ is the wave vector ($V_\textup{\scriptsize{b}}$ is the drift velocity of the ion beam), $\lambda _\textup{\scriptsize{De}}$ is the Debye length, and $n$ denotes the $n$th harmonic. By plugging the observed values in the above equation, we find that the maximum growth rate is about 130 $\textup{s}^{-1}$ or growth time $8\times 10^{-3}$ s. In contrast, the time for solar wind electrons to travel a distance of 50 km is about  $10^{-1}$ s, much longer than the wave growth time. Therefore, ECDI is very capable of driving the perpendicular electrostatic fluctuations observed in Figure~\ref{fig:case14}g.


\subsection{Discussion}
\label{subsec:discussion}

As discussed earlier, ECDI can result in electrostatic waves propagating perpendicular to the magnetic field. However, so far we have not considered the effect of the electron parallel motion in the dispersion relation of the ECDI. If we allow a finite wave vector along the magnetic field in the dispersion relation, then a new type of instability naturally arises in the solution: modified two-stream instability (MTSI) \cite{janhunen_evolution_2018}. Due to the parallel component of the wave fields, previous studies have shown that the MTSI can cause strong parallel heating of electrons \cite{mcbride_theory_1972, matsukiyo_microinstabilities_2006}. In section~\ref{sec:heating}, we will show that ECDI, along with MTSI, may provide a mechanism responsible for substantial isotropic electron heating as observed over lunar magnetic anomalies.



\section{Wave-Particle Interaction and Electron Heating}
\label{sec:heating}

When waves are traveling through a plasma, the fluctuating wave fields can interact with the charged particles of the plasma, resulting in many interesting nonlinear effects. As shown in section~\ref{sec:observations}, one of the important features we notice in the two flyby events is the significant electron heating observed over lunar crustal magnetic anomalies (Figures~\ref{fig:case14}f and \ref{fig:case5}f). Furthermore, the electric and magnetic field FFT wave spectra (Figures~\ref{fig:case14}h--\ref{fig:case14}i and \ref{fig:case5}h--\ref{fig:case5}i) reveal that the electron heating is always accompanied by significant electrostatic and/or electromagnetic wave power, suggesting that the wave-particle interaction may play an important role in heating the electrons.

\begin{figure*}
\begin{center}
\includegraphics[width=6.69in]{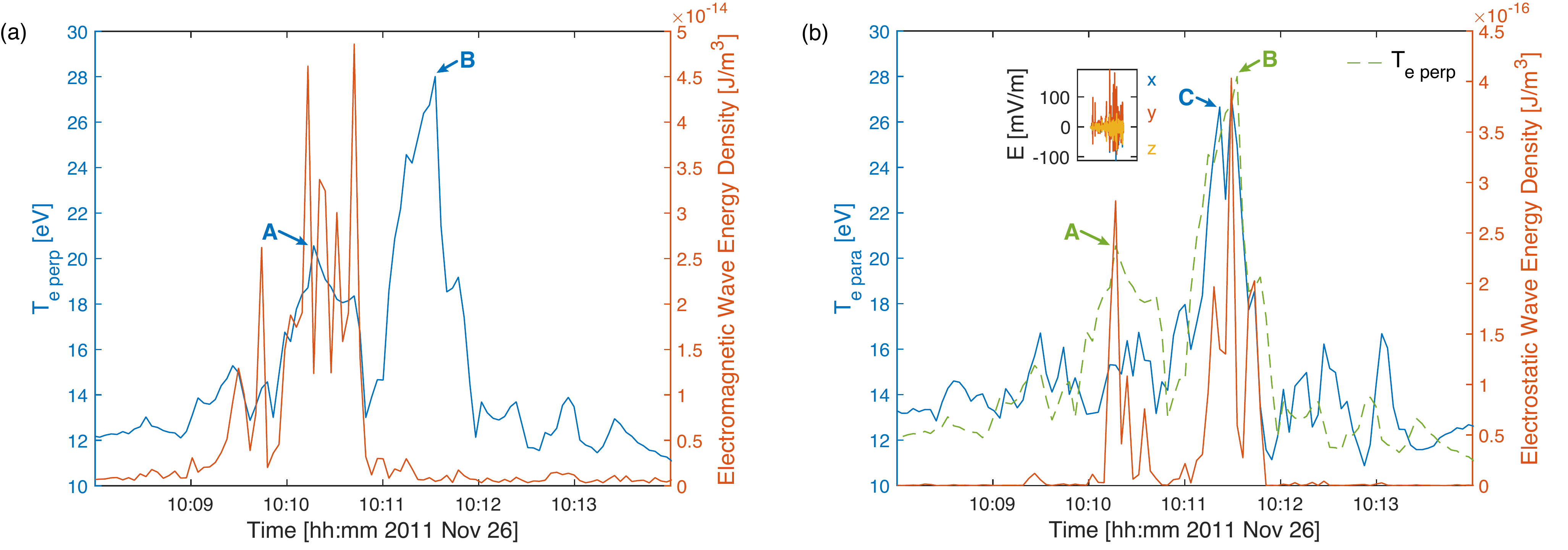}
\caption{(a) Perpendicular electron temperature and electromagnetic wave energy density (frequencies ranging from near-DC levels to tens of Hz) as a function of time for the P2 lunar flyby event shown in Figure~\ref{fig:case14}. (b) Parallel electron temperature and electrostatic wave energy density (frequencies ranging from 10 Hz to 8 kHz) as a function of time for the same flyby. The perpendicular temperature is also shown in the background for comparison. Inset shows the same electric field as Figure~\ref{fig:case14}e. The largest two peaks in the perpendicular temperature are denoted by A and B, respectively. The largest peak in the parallel temperature is denoted by C.}
\label{fig:heating}
\end{center}
\end{figure*}

To demonstrate the mechanisms causing the electron heating above the magnetic anomaly regions, here we only focus on the ARTEMIS P2 lunar flyby as shown in Figure~\ref{fig:case14}. We plot the perpendicular electron temperature and the electromagnetic wave energy density together as a function of time in Figure~\ref{fig:heating}a. We present a similar figure showing the parallel electron temperature and the electrostatic wave energy density as a function of time in Figure~\ref{fig:heating}b, where the perpendicular temperature is also shown for comparison. We notice that there are two peaks (A and B) in the perpendicular electron temperature and one peak (C) in the parallel temperature. Further investigation of the correlation between the wave power and the electron temperature shows that these peaks are caused by different heating mechanisms.


We note that peak A in the perpendicular electron temperature is very well aligned with the electromagnetic wave power in Figure~\ref{fig:heating}a, suggesting that the heating likely results from the cyclotron resonance with the whistler modes. When the electrons encounter the whistler waves Doppler-shifted to their cyclotron frequency or higher harmonics, they can strongly interact with the wave fields, gaining perpendicular energy and causing the waves to damp \cite{tsurutani_basic_1997, stenzel_whistler_2016}. Similar perpendicular heating by electromagnetic waves near the Moon has been reported in \citeA{halekas_lunar_2012} -- though in this case it happens in the magnetotail. In addition to cyclotron damping, we also note that the peak A coincides with the peaks of the electrostatic wave power in Figure~\ref{fig:heating}b, suggesting that the perpendicular electrostatic waves driven by ECDI are likely to be another source contributing to the perpendicular heating in peak A. This electron heating mechanism resulting from ECDI is also observed in the foot region of perpendicular shocks \cite{matsukiyo_microinstabilities_2006}. One can argue that the perpendicular temperature being higher than the parallel temperature could arise from loss cones that are seen in Figure~\ref{fig:case14}d between 10:10 UT and 10:11 UT. However, the parallel temperature is of about the same value even outside the magnetic anomalies, suggesting that the loss cones may not significantly reduce the parallel temperature in our case. In addition, loss cones may lead to a decrease in parallel temperature, but it cannot affect perpendicular temperature. The fact that we observe an increase in the perpendicular temperature shown in Figure~\ref{fig:heating} is therefore caused by electron heating rather than loss cones.

As to peak B, since it is only accompanied by the peaks of the electrostatic wave power in Figure~\ref{fig:heating}b, this heating therefore is likely to be caused by ECDI as well. Last but not the least, peak C may seem to be quite puzzling at first. Even though it is aligned with peak B and accompanied by intense electrostatic wave power, the perpendicular electric fields resulting from ECDI cannot heat the electrons in the parallel direction. However, as discussed in section~\ref{subsec:discussion}, MTSI can be driven unstable in the similar conditions as ECDI. In fact, ECDI and MTSI can often be excited simultaneously, allowing for substantial electron heating both perpendicular and parallel to the magnetic field \cite{wu_microinstabilities_1984, muschietti_microturbulence_2013, janhunen_evolution_2018}. Since ETSI can also lead to parallel heating, therefore, the isotropic heating seen in peaks B and C (Figure~\ref{fig:heating}b) may be caused by a combination of contributions from ECDI, MTSI, and ETSI.


\section{Conclusions}
\label{sec:conclusions}

In conclusion, we have investigated two types of electrostatic instabilities observed over the lunar crustal magnetic anomalies during ARTEMIS flyby. The electrostatic waves propagating parallel to the ambient magnetic field are attributed to upward electron beams reflected by the crustal magnetic fields. We have also reported for the first time on observations of another class of electrostatic waves propagating perpendicular to the magnetic field. A proposed free‐energy source is associated with reflected ion beams streaming across the background magnetic field. Finally, our analysis suggests that the perpendicular electron heating observed above the magnetic anomalies is mainly caused by cyclotron damping and ECDI. The isotropic heating, on the other hand, may result from joint effects due to ECDI, MTSI, and ETSI.

\acknowledgments
We acknowledge support from Solar System Exploration Research Virtual Institute and the THEMIS Contract NAS5‐02099. K.H.G. is financially supported by the German Ministerium für Wirtschaft und Energie and the Deutsches Zentrum für Luft und Raumfahrt under contract 50 OC 1403. All ARTEMIS data used in this paper are publicly available at NASA’s CDAWeb (\url{https://cdaweb.sci.gsfc.nasa.gov}) and the ARTEMIS site (\url{http://artemis.ssl.berkeley.edu}). 



%
%

\bibliography{refs}

\begin{thebibliography}{}

\bibitem [\protect \citeauthoryear {%
Andersson%
\ \protect \BOthers {.}}{%
Andersson%
\ \protect \BOthers {.}}{%
{\protect \APACyear {2002}}%
}]{%
andersson_characteristics_2002}
\APACinsertmetastar {%
andersson_characteristics_2002}%
\begin{APACrefauthors}%
Andersson, L.%
, Ergun, R\BPBI E.%
, Newman, D\BPBI L.%
, McFadden, J\BPBI P.%
, Carlson, C\BPBI W.%
\BCBL {}\ \BBA {} Su, Y\BHBI J.%
\end{APACrefauthors}%
\unskip\
\newblock
\APACrefYearMonthDay{2002}{{\APACmonth{07}}}{}.
\newblock
{\BBOQ}\APACrefatitle {Characteristics of Parallel Electric Fields in the
  Downward Current Region of the Aurora} {Characteristics of parallel electric
  fields in the downward current region of the aurora}.{\BBCQ}
\newblock
\APACjournalVolNumPages{Physics of Plasmas}{9}{8}{3600--3609}.
\newblock
\begin{APACrefDOI} \doi{10.1063/1.1490134} \end{APACrefDOI}
\PrintBackRefs{\CurrentBib}

\bibitem [\protect \citeauthoryear {%
Angelopoulos%
}{%
Angelopoulos%
}{%
{\protect \APACyear {2011}}%
}]{%
angelopoulos_artemis_2011}
\APACinsertmetastar {%
angelopoulos_artemis_2011}%
\begin{APACrefauthors}%
Angelopoulos, V.%
\end{APACrefauthors}%
\unskip\
\newblock
\APACrefYearMonthDay{2011}{{\APACmonth{12}}}{}.
\newblock
{\BBOQ}\APACrefatitle {The {{ARTEMIS Mission}}} {The {{ARTEMIS
  Mission}}}.{\BBCQ}
\newblock
\APACjournalVolNumPages{Space Sci Rev}{165}{1}{3--25}.
\newblock
\begin{APACrefDOI} \doi{10.1007/s11214-010-9687-2} \end{APACrefDOI}
\PrintBackRefs{\CurrentBib}

\bibitem [\protect \citeauthoryear {%
Auster%
\ \protect \BOthers {.}}{%
Auster%
\ \protect \BOthers {.}}{%
{\protect \APACyear {2008}}%
}]{%
auster_themis_2008}
\APACinsertmetastar {%
auster_themis_2008}%
\begin{APACrefauthors}%
Auster, H\BPBI U.%
, Glassmeier, K\BPBI H.%
, Magnes, W.%
, Aydogar, O.%
, Baumjohann, W.%
, Constantinescu, D.%
\BDBL {}Wiedemann, M.%
\end{APACrefauthors}%
\unskip\
\newblock
\APACrefYearMonthDay{2008}{{\APACmonth{12}}}{}.
\newblock
{\BBOQ}\APACrefatitle {The {{THEMIS Fluxgate Magnetometer}}} {The {{THEMIS
  Fluxgate Magnetometer}}}.{\BBCQ}
\newblock
\APACjournalVolNumPages{Space Sci Rev}{141}{1}{235--264}.
\newblock
\begin{APACrefDOI} \doi{10.1007/s11214-008-9365-9} \end{APACrefDOI}
\PrintBackRefs{\CurrentBib}

\bibitem [\protect \citeauthoryear {%
Bale%
\ \protect \BOthers {.}}{%
Bale%
\ \protect \BOthers {.}}{%
{\protect \APACyear {1998}}%
}]{%
bale_bipolar_1998}
\APACinsertmetastar {%
bale_bipolar_1998}%
\begin{APACrefauthors}%
Bale, S\BPBI D.%
, Kellogg, P\BPBI J.%
, Larsen, D\BPBI E.%
, Lin, R\BPBI P.%
, Goetz, K.%
\BCBL {}\ \BBA {} Lepping, R\BPBI P.%
\end{APACrefauthors}%
\unskip\
\newblock
\APACrefYearMonthDay{1998}{}{}.
\newblock
{\BBOQ}\APACrefatitle {Bipolar Electrostatic Structures in the Shock Transition
  Region: {{Evidence}} of Electron Phase Space Holes} {Bipolar electrostatic
  structures in the shock transition region: {{Evidence}} of electron phase
  space holes}.{\BBCQ}
\newblock
\APACjournalVolNumPages{Geophys. Res. Lett.}{25}{15}{2929--2932}.
\newblock
\begin{APACrefDOI} \doi{10.1029/98GL02111} \end{APACrefDOI}
\PrintBackRefs{\CurrentBib}

\bibitem [\protect \citeauthoryear {%
Bhardwaj%
\ \protect \BOthers {.}}{%
Bhardwaj%
\ \protect \BOthers {.}}{%
{\protect \APACyear {2015}}%
}]{%
bhardwaj_new_2015}
\APACinsertmetastar {%
bhardwaj_new_2015}%
\begin{APACrefauthors}%
Bhardwaj, A.%
, Dhanya, M\BPBI B.%
, Alok, A.%
, Barabash, S.%
, Wieser, M.%
, Futaana, Y.%
\BDBL {}Asamura, K.%
\end{APACrefauthors}%
\unskip\
\newblock
\APACrefYearMonthDay{2015}{{\APACmonth{08}}}{}.
\newblock
{\BBOQ}\APACrefatitle {A New View on the Solar Wind Interaction with the
  {{Moon}}} {A new view on the solar wind interaction with the
  {{Moon}}}.{\BBCQ}
\newblock
\APACjournalVolNumPages{Geoscience Letters}{2}{1}{10}.
\newblock
\begin{APACrefDOI} \doi{10.1186/s40562-015-0027-y} \end{APACrefDOI}
\PrintBackRefs{\CurrentBib}

\bibitem [\protect \citeauthoryear {%
Bonnell%
\ \protect \BOthers {.}}{%
Bonnell%
\ \protect \BOthers {.}}{%
{\protect \APACyear {2008}}%
}]{%
bonnell_electric_2008}
\APACinsertmetastar {%
bonnell_electric_2008}%
\begin{APACrefauthors}%
Bonnell, J\BPBI W.%
, Mozer, F\BPBI S.%
, Delory, G\BPBI T.%
, Hull, A\BPBI J.%
, Ergun, R\BPBI E.%
, Cully, C\BPBI M.%
\BDBL {}Harvey, P\BPBI R.%
\end{APACrefauthors}%
\unskip\
\newblock
\APACrefYearMonthDay{2008}{{\APACmonth{12}}}{}.
\newblock
{\BBOQ}\APACrefatitle {The {{Electric Field Instrument}} ({{EFI}}) for
  {{THEMIS}}} {The {{Electric Field Instrument}} ({{EFI}}) for
  {{THEMIS}}}.{\BBCQ}
\newblock
\APACjournalVolNumPages{Space Sci Rev}{141}{1}{303--341}.
\newblock
\begin{APACrefDOI} \doi{10.1007/s11214-008-9469-2} \end{APACrefDOI}
\PrintBackRefs{\CurrentBib}

\bibitem [\protect \citeauthoryear {%
Cao%
, Halekas%
, Poppe%
, Chu%
\BCBL {}\ \BBA {} Glassmeier%
}{%
Cao%
, Halekas%
, Poppe%
\BCBL {}\ \protect \BOthers {.}}{%
{\protect \APACyear {2020}}%
}]{%
cao_acceleration_2020}
\APACinsertmetastar {%
cao_acceleration_2020}%
\begin{APACrefauthors}%
Cao, X.%
, Halekas, J.%
, Poppe, A.%
, Chu, F.%
\BCBL {}\ \BBA {} Glassmeier, K\BHBI H.%
\end{APACrefauthors}%
\unskip\
\newblock
\APACrefYearMonthDay{2020}{}{}.
\newblock
{\BBOQ}\APACrefatitle {The {{Acceleration}} of {{Lunar Ions}} by {{Magnetic
  Forces}} in the {{Terrestrial Magnetotail Lobes}}} {The {{Acceleration}} of
  {{Lunar Ions}} by {{Magnetic Forces}} in the {{Terrestrial Magnetotail
  Lobes}}}.{\BBCQ}
\newblock
\APACjournalVolNumPages{J. Geophys. Res. Space Phys.}{125}{6}{e2020JA027829}.
\newblock
\begin{APACrefDOI} \doi{10.1029/2020JA027829} \end{APACrefDOI}
\PrintBackRefs{\CurrentBib}

\bibitem [\protect \citeauthoryear {%
Cao%
, Halekas%
, Chu%
\BCBL {}\ \protect \BOthers {.}}{%
Cao%
, Halekas%
, Chu%
\BCBL {}\ \protect \BOthers {.}}{%
{\protect \APACyear {2020}}%
}]{%
cao_plasma_2020}
\APACinsertmetastar {%
cao_plasma_2020}%
\begin{APACrefauthors}%
Cao, X.%
, Halekas, J\BPBI S.%
, Chu, F.%
, Kistler, M.%
, Poppe, A\BPBI R.%
\BCBL {}\ \BBA {} Glassmeier, K\BHBI H.%
\end{APACrefauthors}%
\unskip\
\newblock
\APACrefYearMonthDay{2020}{{\APACmonth{10}}}{}.
\newblock
{\BBOQ}\APACrefatitle {Plasma {{Convection}} in the {{Terrestrial Magnetotail
  Lobes Measured Near}} the {{Moon}}'s {{Orbit}}} {Plasma {{Convection}} in the
  {{Terrestrial Magnetotail Lobes Measured Near}} the {{Moon}}'s
  {{Orbit}}}.{\BBCQ}
\newblock
\APACjournalVolNumPages{Geophys. Res. Lett.}{47}{20}{}.
\newblock
\begin{APACrefDOI} \doi{10.1029/2020GL090217} \end{APACrefDOI}
\PrintBackRefs{\CurrentBib}

\bibitem [\protect \citeauthoryear {%
Ergun%
, Andersson%
, Carlson%
, Newman%
\BCBL {}\ \BBA {} Goldman%
}{%
Ergun%
\ \protect \BOthers {.}}{%
{\protect \APACyear {2003}}%
}]{%
ergun_double_2003}
\APACinsertmetastar {%
ergun_double_2003}%
\begin{APACrefauthors}%
Ergun, R\BPBI E.%
, Andersson, L.%
, Carlson, C\BPBI W.%
, Newman, D\BPBI L.%
\BCBL {}\ \BBA {} Goldman, M\BPBI V.%
\end{APACrefauthors}%
\unskip\
\newblock
\APACrefYearMonthDay{2003}{}{}.
\newblock
{\BBOQ}\APACrefatitle {Double Layers in the Downward Current Region of the
  Aurora} {Double layers in the downward current region of the aurora}.{\BBCQ}
\newblock
\APACjournalVolNumPages{Nonlinear Process. Geophys.}{10}{1/2}{45--52}.
\PrintBackRefs{\CurrentBib}

\bibitem [\protect \citeauthoryear {%
Forslund%
, Morse%
\BCBL {}\ \BBA {} Nielson%
}{%
Forslund%
\ \protect \BOthers {.}}{%
{\protect \APACyear {1970}}%
}]{%
forslund_electron_1970}
\APACinsertmetastar {%
forslund_electron_1970}%
\begin{APACrefauthors}%
Forslund, D\BPBI W.%
, Morse, R\BPBI L.%
\BCBL {}\ \BBA {} Nielson, C\BPBI W.%
\end{APACrefauthors}%
\unskip\
\newblock
\APACrefYearMonthDay{1970}{{\APACmonth{11}}}{}.
\newblock
{\BBOQ}\APACrefatitle {Electron {{Cyclotron Drift Instability}}} {Electron
  {{Cyclotron Drift Instability}}}.{\BBCQ}
\newblock
\APACjournalVolNumPages{Phys. Rev. Lett.}{25}{18}{1266--1270}.
\newblock
\begin{APACrefDOI} \doi{10.1103/PhysRevLett.25.1266} \end{APACrefDOI}
\PrintBackRefs{\CurrentBib}

\bibitem [\protect \citeauthoryear {%
Franz%
, Kintner%
, Pickett%
\BCBL {}\ \BBA {} Chen%
}{%
Franz%
\ \protect \BOthers {.}}{%
{\protect \APACyear {2005}}%
}]{%
franz_properties_2005}
\APACinsertmetastar {%
franz_properties_2005}%
\begin{APACrefauthors}%
Franz, J\BPBI R.%
, Kintner, P\BPBI M.%
, Pickett, J\BPBI S.%
\BCBL {}\ \BBA {} Chen, L\BHBI J.%
\end{APACrefauthors}%
\unskip\
\newblock
\APACrefYearMonthDay{2005}{}{}.
\newblock
{\BBOQ}\APACrefatitle {Properties of Small-Amplitude Electron Phase-Space Holes
  Observed by {{Polar}}} {Properties of small-amplitude electron phase-space
  holes observed by {{Polar}}}.{\BBCQ}
\newblock
\APACjournalVolNumPages{J. Geophys. Res. Space Phys.}{110}{A9}{}.
\newblock
\begin{APACrefDOI} \doi{10.1029/2005JA011095} \end{APACrefDOI}
\PrintBackRefs{\CurrentBib}

\bibitem [\protect \citeauthoryear {%
Goldstein%
}{%
Goldstein%
}{%
{\protect \APACyear {1974}}%
}]{%
goldstein_observations_1974}
\APACinsertmetastar {%
goldstein_observations_1974}%
\begin{APACrefauthors}%
Goldstein, B\BPBI E.%
\end{APACrefauthors}%
\unskip\
\newblock
\APACrefYearMonthDay{1974}{}{}.
\newblock
{\BBOQ}\APACrefatitle {Observations of Electrons at the Lunar Surface}
  {Observations of electrons at the lunar surface}.{\BBCQ}
\newblock
\APACjournalVolNumPages{J. Geophys. Res. 1896-1977}{79}{1}{23--35}.
\newblock
\begin{APACrefDOI} \doi{10.1029/JA079i001p00023} \end{APACrefDOI}
\PrintBackRefs{\CurrentBib}

\bibitem [\protect \citeauthoryear {%
Graham%
, Khotyaintsev%
, Vaivads%
\BCBL {}\ \BBA {} Andr{\'e}%
}{%
Graham%
\ \protect \BOthers {.}}{%
{\protect \APACyear {2016}}%
}]{%
graham_electrostatic_2016}
\APACinsertmetastar {%
graham_electrostatic_2016}%
\begin{APACrefauthors}%
Graham, D\BPBI B.%
, Khotyaintsev, Y\BPBI V.%
, Vaivads, A.%
\BCBL {}\ \BBA {} Andr{\'e}, M.%
\end{APACrefauthors}%
\unskip\
\newblock
\APACrefYearMonthDay{2016}{}{}.
\newblock
{\BBOQ}\APACrefatitle {Electrostatic Solitary Waves and Electrostatic Waves at
  the Magnetopause} {Electrostatic solitary waves and electrostatic waves at
  the magnetopause}.{\BBCQ}
\newblock
\APACjournalVolNumPages{J. Geophys. Res. Space Phys.}{121}{4}{3069--3092}.
\newblock
\begin{APACrefDOI} \doi{10.1002/2015JA021527} \end{APACrefDOI}
\PrintBackRefs{\CurrentBib}

\bibitem [\protect \citeauthoryear {%
J.~Halekas%
, Poppe%
, Delory%
, Farrell%
\BCBL {}\ \BBA {} Hor{\'a}nyi%
}{%
J.~Halekas%
\ \protect \BOthers {.}}{%
{\protect \APACyear {2012}}%
}]{%
halekas_solar_2012}
\APACinsertmetastar {%
halekas_solar_2012}%
\begin{APACrefauthors}%
Halekas, J.%
, Poppe, A.%
, Delory, G.%
, Farrell, W.%
\BCBL {}\ \BBA {} Hor{\'a}nyi, M.%
\end{APACrefauthors}%
\unskip\
\newblock
\APACrefYearMonthDay{2012}{{\APACmonth{02}}}{}.
\newblock
{\BBOQ}\APACrefatitle {Solar Wind Electron Interaction with the Dayside Lunar
  Surface and Crustal Magnetic Fields: {{Evidence}} for Precursor Effects}
  {Solar wind electron interaction with the dayside lunar surface and crustal
  magnetic fields: {{Evidence}} for precursor effects}.{\BBCQ}
\newblock
\APACjournalVolNumPages{Earth Planets Space}{64}{2}{73--82}.
\newblock
\begin{APACrefDOI} \doi{10.5047/eps.2011.03.008} \end{APACrefDOI}
\PrintBackRefs{\CurrentBib}

\bibitem [\protect \citeauthoryear {%
J\BPBI S.~Halekas%
\ \protect \BOthers {.}}{%
J\BPBI S.~Halekas%
\ \protect \BOthers {.}}{%
{\protect \APACyear {2001}}%
}]{%
halekas_mapping_2001}
\APACinsertmetastar {%
halekas_mapping_2001}%
\begin{APACrefauthors}%
Halekas, J\BPBI S.%
, Mitchell, D\BPBI L.%
, Lin, R\BPBI P.%
, Frey, S.%
, Hood, L\BPBI L.%
, Acu{\~n}a, M\BPBI H.%
\BCBL {}\ \BBA {} Binder, A\BPBI B.%
\end{APACrefauthors}%
\unskip\
\newblock
\APACrefYearMonthDay{2001}{}{}.
\newblock
{\BBOQ}\APACrefatitle {Mapping of Crustal Magnetic Anomalies on the Lunar near
  Side by the {{Lunar Prospector}} Electron Reflectometer} {Mapping of crustal
  magnetic anomalies on the lunar near side by the {{Lunar Prospector}}
  electron reflectometer}.{\BBCQ}
\newblock
\APACjournalVolNumPages{J. Geophys. Res. Planets}{106}{E11}{27841--27852}.
\newblock
\begin{APACrefDOI} \doi{10.1029/2000JE001380} \end{APACrefDOI}
\PrintBackRefs{\CurrentBib}

\bibitem [\protect \citeauthoryear {%
J\BPBI S.~Halekas%
\ \protect \BOthers {.}}{%
J\BPBI S.~Halekas%
\ \protect \BOthers {.}}{%
{\protect \APACyear {2002}}%
}]{%
halekas_evidence_2002}
\APACinsertmetastar {%
halekas_evidence_2002}%
\begin{APACrefauthors}%
Halekas, J\BPBI S.%
, Mitchell, D\BPBI L.%
, Lin, R\BPBI P.%
, Hood, L\BPBI L.%
, Acu{\~n}a, M\BPBI H.%
\BCBL {}\ \BBA {} Binder, A\BPBI B.%
\end{APACrefauthors}%
\unskip\
\newblock
\APACrefYearMonthDay{2002}{}{}.
\newblock
{\BBOQ}\APACrefatitle {Evidence for Negative Charging of the Lunar Surface in
  Shadow} {Evidence for negative charging of the lunar surface in
  shadow}.{\BBCQ}
\newblock
\APACjournalVolNumPages{Geophys. Res. Lett.}{29}{10}{77-1-77-4}.
\newblock
\begin{APACrefDOI} \doi{10.1029/2001GL014428} \end{APACrefDOI}
\PrintBackRefs{\CurrentBib}

\bibitem [\protect \citeauthoryear {%
J\BPBI S.~Halekas%
, Poppe%
, Delory%
\BCBL {}\ \protect \BOthers {.}}{%
J\BPBI S.~Halekas%
, Poppe%
, Delory%
\BCBL {}\ \protect \BOthers {.}}{%
{\protect \APACyear {2012}}%
}]{%
halekas_lunar_2012-1}
\APACinsertmetastar {%
halekas_lunar_2012-1}%
\begin{APACrefauthors}%
Halekas, J\BPBI S.%
, Poppe, A\BPBI R.%
, Delory, G\BPBI T.%
, Sarantos, M.%
, Farrell, W\BPBI M.%
, Angelopoulos, V.%
\BCBL {}\ \BBA {} McFadden, J\BPBI P.%
\end{APACrefauthors}%
\unskip\
\newblock
\APACrefYearMonthDay{2012}{}{}.
\newblock
{\BBOQ}\APACrefatitle {Lunar Pickup Ions Observed by {{ARTEMIS}}: {{Spatial}}
  and Temporal Distribution and Constraints on Species and Source Locations}
  {Lunar pickup ions observed by {{ARTEMIS}}: {{Spatial}} and temporal
  distribution and constraints on species and source locations}.{\BBCQ}
\newblock
\APACjournalVolNumPages{J. Geophys. Res. Planets}{117}{E6}{}.
\newblock
\begin{APACrefDOI} \doi{10.1029/2012JE004107} \end{APACrefDOI}
\PrintBackRefs{\CurrentBib}

\bibitem [\protect \citeauthoryear {%
J\BPBI S.~Halekas%
, Poppe%
, Farrell%
\BCBL {}\ \protect \BOthers {.}}{%
J\BPBI S.~Halekas%
, Poppe%
, Farrell%
\BCBL {}\ \protect \BOthers {.}}{%
{\protect \APACyear {2012}}%
}]{%
halekas_lunar_2012}
\APACinsertmetastar {%
halekas_lunar_2012}%
\begin{APACrefauthors}%
Halekas, J\BPBI S.%
, Poppe, A\BPBI R.%
, Farrell, W\BPBI M.%
, Delory, G\BPBI T.%
, Angelopoulos, V.%
, McFadden, J\BPBI P.%
\BDBL {}Ergun, R\BPBI E.%
\end{APACrefauthors}%
\unskip\
\newblock
\APACrefYearMonthDay{2012}{}{}.
\newblock
{\BBOQ}\APACrefatitle {Lunar Precursor Effects in the Solar Wind and
  Terrestrial Magnetosphere} {Lunar precursor effects in the solar wind and
  terrestrial magnetosphere}.{\BBCQ}
\newblock
\APACjournalVolNumPages{J. Geophys. Res. Space Phys.}{117}{A5}{}.
\newblock
\begin{APACrefDOI} \doi{10.1029/2011JA017289} \end{APACrefDOI}
\PrintBackRefs{\CurrentBib}

\bibitem [\protect \citeauthoryear {%
J\BPBI S.~Halekas%
\ \protect \BOthers {.}}{%
J\BPBI S.~Halekas%
\ \protect \BOthers {.}}{%
{\protect \APACyear {2014}}%
}]{%
halekas_evidence_2014}
\APACinsertmetastar {%
halekas_evidence_2014}%
\begin{APACrefauthors}%
Halekas, J\BPBI S.%
, Poppe, A\BPBI R.%
, McFadden, J\BPBI P.%
, Angelopoulos, V.%
, Glassmeier, K\BHBI H.%
\BCBL {}\ \BBA {} Brain, D\BPBI A.%
\end{APACrefauthors}%
\unskip\
\newblock
\APACrefYearMonthDay{2014}{}{}.
\newblock
{\BBOQ}\APACrefatitle {Evidence for Small-Scale Collisionless Shocks at the
  {{Moon}} from {{ARTEMIS}}} {Evidence for small-scale collisionless shocks at
  the {{Moon}} from {{ARTEMIS}}}.{\BBCQ}
\newblock
\APACjournalVolNumPages{Geophys. Res. Lett.}{41}{21}{7436--7443}.
\newblock
\begin{APACrefDOI} \doi{10.1002/2014GL061973} \end{APACrefDOI}
\PrintBackRefs{\CurrentBib}

\bibitem [\protect \citeauthoryear {%
J\BPBI S.~Halekas%
, Saito%
, Delory%
\BCBL {}\ \BBA {} Farrell%
}{%
J\BPBI S.~Halekas%
\ \protect \BOthers {.}}{%
{\protect \APACyear {2011}}%
}]{%
halekas_new_2011}
\APACinsertmetastar {%
halekas_new_2011}%
\begin{APACrefauthors}%
Halekas, J\BPBI S.%
, Saito, Y.%
, Delory, G\BPBI T.%
\BCBL {}\ \BBA {} Farrell, W\BPBI M.%
\end{APACrefauthors}%
\unskip\
\newblock
\APACrefYearMonthDay{2011}{{\APACmonth{11}}}{}.
\newblock
{\BBOQ}\APACrefatitle {New Views of the Lunar Plasma Environment} {New views of
  the lunar plasma environment}.{\BBCQ}
\newblock
\APACjournalVolNumPages{Planetary and Space Science}{59}{14}{1681--1694}.
\newblock
\begin{APACrefDOI} \doi{10.1016/j.pss.2010.08.011} \end{APACrefDOI}
\PrintBackRefs{\CurrentBib}

\bibitem [\protect \citeauthoryear {%
Harada%
\ \BBA {} Halekas%
}{%
Harada%
\ \BBA {} Halekas%
}{%
{\protect \APACyear {2016}}%
}]{%
harada_upstream_2016}
\APACinsertmetastar {%
harada_upstream_2016}%
\begin{APACrefauthors}%
Harada, Y.%
\BCBT {}\ \BBA {} Halekas, J\BPBI S.%
\end{APACrefauthors}%
\unskip\
\newblock
\APACrefYearMonthDay{2016}{}{}.
\newblock
{\BBOQ}\APACrefatitle {Upstream {{Waves}} and {{Particles}} at the {{Moon}}}
  {Upstream {{Waves}} and {{Particles}} at the {{Moon}}}.{\BBCQ}
\newblock
\BIn{} \APACrefbtitle {Low-{{Frequency Waves}} in {{Space Plasmas}}}
  {Low-{{Frequency Waves}} in {{Space Plasmas}}}\ (\BPGS\ 307--322).
\newblock
\APACaddressPublisher{}{{American Geophysical Union (AGU)}}.
\newblock
\begin{APACrefDOI} \doi{10.1002/9781119055006.ch18} \end{APACrefDOI}
\PrintBackRefs{\CurrentBib}

\bibitem [\protect \citeauthoryear {%
Harada%
, Halekas%
, Poppe%
, Kurita%
\BCBL {}\ \BBA {} McFadden%
}{%
Harada%
\ \protect \BOthers {.}}{%
{\protect \APACyear {2014}}%
}]{%
harada_extended_2014}
\APACinsertmetastar {%
harada_extended_2014}%
\begin{APACrefauthors}%
Harada, Y.%
, Halekas, J\BPBI S.%
, Poppe, A\BPBI R.%
, Kurita, S.%
\BCBL {}\ \BBA {} McFadden, J\BPBI P.%
\end{APACrefauthors}%
\unskip\
\newblock
\APACrefYearMonthDay{2014}{}{}.
\newblock
{\BBOQ}\APACrefatitle {Extended Lunar Precursor Regions: {{Electron}}-Wave
  Interaction} {Extended lunar precursor regions: {{Electron}}-wave
  interaction}.{\BBCQ}
\newblock
\APACjournalVolNumPages{J. Geophys. Res. Space Phys.}{119}{11}{9160--9173}.
\newblock
\begin{APACrefDOI} \doi{10.1002/2014JA020618} \end{APACrefDOI}
\PrintBackRefs{\CurrentBib}

\bibitem [\protect \citeauthoryear {%
Harada%
, Poppe%
, Halekas%
, Chamberlin%
\BCBL {}\ \BBA {} McFadden%
}{%
Harada%
\ \protect \BOthers {.}}{%
{\protect \APACyear {2017}}%
}]{%
harada_photoemission_2017}
\APACinsertmetastar {%
harada_photoemission_2017}%
\begin{APACrefauthors}%
Harada, Y.%
, Poppe, A\BPBI R.%
, Halekas, J\BPBI S.%
, Chamberlin, P\BPBI C.%
\BCBL {}\ \BBA {} McFadden, J\BPBI P.%
\end{APACrefauthors}%
\unskip\
\newblock
\APACrefYearMonthDay{2017}{}{}.
\newblock
{\BBOQ}\APACrefatitle {Photoemission and Electrostatic Potentials on the
  Dayside Lunar Surface in the Terrestrial Magnetotail Lobes} {Photoemission
  and electrostatic potentials on the dayside lunar surface in the terrestrial
  magnetotail lobes}.{\BBCQ}
\newblock
\APACjournalVolNumPages{Geophys. Res. Lett.}{44}{11}{5276--5282}.
\newblock
\begin{APACrefDOI} \doi{10.1002/2017GL073419} \end{APACrefDOI}
\PrintBackRefs{\CurrentBib}

\bibitem [\protect \citeauthoryear {%
Hashimoto%
\ \protect \BOthers {.}}{%
Hashimoto%
\ \protect \BOthers {.}}{%
{\protect \APACyear {2010}}%
}]{%
hashimoto_electrostatic_2010}
\APACinsertmetastar {%
hashimoto_electrostatic_2010}%
\begin{APACrefauthors}%
Hashimoto, K.%
, Hashitani, M.%
, Kasahara, Y.%
, Omura, Y.%
, Nishino, M\BPBI N.%
, Saito, Y.%
\BDBL {}Takahashi, F.%
\end{APACrefauthors}%
\unskip\
\newblock
\APACrefYearMonthDay{2010}{}{}.
\newblock
{\BBOQ}\APACrefatitle {Electrostatic Solitary Waves Associated with Magnetic
  Anomalies and Wake Boundary of the {{Moon}} Observed by {{KAGUYA}}}
  {Electrostatic solitary waves associated with magnetic anomalies and wake
  boundary of the {{Moon}} observed by {{KAGUYA}}}.{\BBCQ}
\newblock
\APACjournalVolNumPages{Geophys. Res. Lett.}{37}{19}{}.
\newblock
\begin{APACrefDOI} \doi{10.1029/2010GL044529} \end{APACrefDOI}
\PrintBackRefs{\CurrentBib}

\bibitem [\protect \citeauthoryear {%
Holmes%
\ \protect \BOthers {.}}{%
Holmes%
\ \protect \BOthers {.}}{%
{\protect \APACyear {2018}}%
}]{%
holmes_electron_2018}
\APACinsertmetastar {%
holmes_electron_2018}%
\begin{APACrefauthors}%
Holmes, J\BPBI C.%
, Ergun, R\BPBI E.%
, Newman, D\BPBI L.%
, Ahmadi, N.%
, Andersson, L.%
, Contel, O\BPBI L.%
\BDBL {}Burch, J\BPBI L.%
\end{APACrefauthors}%
\unskip\
\newblock
\APACrefYearMonthDay{2018}{}{}.
\newblock
{\BBOQ}\APACrefatitle {Electron {{Phase}}-{{Space Holes}} in {{Three
  Dimensions}}: {{Multispacecraft Observations}} by {{Magnetospheric
  Multiscale}}} {Electron {{Phase}}-{{Space Holes}} in {{Three Dimensions}}:
  {{Multispacecraft Observations}} by {{Magnetospheric Multiscale}}}.{\BBCQ}
\newblock
\APACjournalVolNumPages{J. Geophys. Res. Space Phys.}{123}{12}{9963--9978}.
\newblock
\begin{APACrefDOI} \doi{10.1029/2018JA025750} \end{APACrefDOI}
\PrintBackRefs{\CurrentBib}

\bibitem [\protect \citeauthoryear {%
Hutchinson%
}{%
Hutchinson%
}{%
{\protect \APACyear {2017}}%
}]{%
hutchinson_electron_2017}
\APACinsertmetastar {%
hutchinson_electron_2017}%
\begin{APACrefauthors}%
Hutchinson, I\BPBI H.%
\end{APACrefauthors}%
\unskip\
\newblock
\APACrefYearMonthDay{2017}{{\APACmonth{03}}}{}.
\newblock
{\BBOQ}\APACrefatitle {Electron Holes in Phase Space: {{What}} They Are and Why
  They Matter} {Electron holes in phase space: {{What}} they are and why they
  matter}.{\BBCQ}
\newblock
\APACjournalVolNumPages{Physics of Plasmas}{24}{5}{055601}.
\newblock
\begin{APACrefDOI} \doi{10.1063/1.4976854} \end{APACrefDOI}
\PrintBackRefs{\CurrentBib}

\bibitem [\protect \citeauthoryear {%
Janhunen%
\ \protect \BOthers {.}}{%
Janhunen%
\ \protect \BOthers {.}}{%
{\protect \APACyear {2018}}%
}]{%
janhunen_evolution_2018}
\APACinsertmetastar {%
janhunen_evolution_2018}%
\begin{APACrefauthors}%
Janhunen, S.%
, Smolyakov, A.%
, Sydorenko, D.%
, Jimenez, M.%
, Kaganovich, I.%
\BCBL {}\ \BBA {} Raitses, Y.%
\end{APACrefauthors}%
\unskip\
\newblock
\APACrefYearMonthDay{2018}{{\APACmonth{08}}}{}.
\newblock
{\BBOQ}\APACrefatitle {Evolution of the Electron Cyclotron Drift Instability in
  Two-Dimensions} {Evolution of the electron cyclotron drift instability in
  two-dimensions}.{\BBCQ}
\newblock
\APACjournalVolNumPages{Physics of Plasmas}{25}{8}{082308}.
\newblock
\begin{APACrefDOI} \doi{10.1063/1.5033896} \end{APACrefDOI}
\PrintBackRefs{\CurrentBib}

\bibitem [\protect \citeauthoryear {%
Jao%
\ \BBA {} Hau%
}{%
Jao%
\ \BBA {} Hau%
}{%
{\protect \APACyear {2014}}%
}]{%
jao_fluid_2014}
\APACinsertmetastar {%
jao_fluid_2014}%
\begin{APACrefauthors}%
Jao, C\BHBI S.%
\BCBT {}\ \BBA {} Hau, L\BHBI N.%
\end{APACrefauthors}%
\unskip\
\newblock
\APACrefYearMonthDay{2014}{{\APACmonth{02}}}{}.
\newblock
{\BBOQ}\APACrefatitle {Fluid Aspects of Electron Streaming Instability in
  Electron-Ion Plasmas} {Fluid aspects of electron streaming instability in
  electron-ion plasmas}.{\BBCQ}
\newblock
\APACjournalVolNumPages{Physics of Plasmas}{21}{2}{022103}.
\newblock
\begin{APACrefDOI} \doi{10.1063/1.4863839} \end{APACrefDOI}
\PrintBackRefs{\CurrentBib}

\bibitem [\protect \citeauthoryear {%
Lue%
\ \protect \BOthers {.}}{%
Lue%
\ \protect \BOthers {.}}{%
{\protect \APACyear {2014}}%
}]{%
lue_chandrayaan-1_2014}
\APACinsertmetastar {%
lue_chandrayaan-1_2014}%
\begin{APACrefauthors}%
Lue, C.%
, Futaana, Y.%
, Barabash, S.%
, Wieser, M.%
, Bhardwaj, A.%
\BCBL {}\ \BBA {} Wurz, P.%
\end{APACrefauthors}%
\unskip\
\newblock
\APACrefYearMonthDay{2014}{}{}.
\newblock
{\BBOQ}\APACrefatitle {Chandrayaan-1 Observations of Backscattered Solar Wind
  Protons from the Lunar Regolith: {{Dependence}} on the Solar Wind Speed}
  {Chandrayaan-1 observations of backscattered solar wind protons from the
  lunar regolith: {{Dependence}} on the solar wind speed}.{\BBCQ}
\newblock
\APACjournalVolNumPages{J. Geophys. Res. Planets}{119}{5}{968--975}.
\newblock
\begin{APACrefDOI} \doi{10.1002/2013JE004582} \end{APACrefDOI}
\PrintBackRefs{\CurrentBib}

\bibitem [\protect \citeauthoryear {%
Lue%
\ \protect \BOthers {.}}{%
Lue%
\ \protect \BOthers {.}}{%
{\protect \APACyear {2011}}%
}]{%
lue_strong_2011}
\APACinsertmetastar {%
lue_strong_2011}%
\begin{APACrefauthors}%
Lue, C.%
, Futaana, Y.%
, Barabash, S.%
, Wieser, M.%
, Holmstr{\"o}m, M.%
, Bhardwaj, A.%
\BDBL {}Wurz, P.%
\end{APACrefauthors}%
\unskip\
\newblock
\APACrefYearMonthDay{2011}{}{}.
\newblock
{\BBOQ}\APACrefatitle {Strong Influence of Lunar Crustal Fields on the Solar
  Wind Flow} {Strong influence of lunar crustal fields on the solar wind
  flow}.{\BBCQ}
\newblock
\APACjournalVolNumPages{Geophys. Res. Lett.}{38}{3}{}.
\newblock
\begin{APACrefDOI} \doi{10.1029/2010GL046215} \end{APACrefDOI}
\PrintBackRefs{\CurrentBib}

\bibitem [\protect \citeauthoryear {%
Malaspina%
\ \protect \BOthers {.}}{%
Malaspina%
\ \protect \BOthers {.}}{%
{\protect \APACyear {2013}}%
}]{%
malaspina_electrostatic_2013}
\APACinsertmetastar {%
malaspina_electrostatic_2013}%
\begin{APACrefauthors}%
Malaspina, D\BPBI M.%
, Newman, D\BPBI L.%
, Willson, L\BPBI B.%
, Goetz, K.%
, Kellogg, P\BPBI J.%
\BCBL {}\ \BBA {} Kerstin, K.%
\end{APACrefauthors}%
\unskip\
\newblock
\APACrefYearMonthDay{2013}{}{}.
\newblock
{\BBOQ}\APACrefatitle {Electrostatic {{Solitary Waves}} in the {{Solar Wind}}:
  {{Evidence}} for {{Instability}} at {{Solar Wind Current Sheets}}}
  {Electrostatic {{Solitary Waves}} in the {{Solar Wind}}: {{Evidence}} for
  {{Instability}} at {{Solar Wind Current Sheets}}}.{\BBCQ}
\newblock
\APACjournalVolNumPages{J. Geophys. Res. Space Phys.}{118}{2}{591--599}.
\newblock
\begin{APACrefDOI} \doi{10.1002/jgra.50102} \end{APACrefDOI}
\PrintBackRefs{\CurrentBib}

\bibitem [\protect \citeauthoryear {%
Matsukiyo%
\ \BBA {} Scholer%
}{%
Matsukiyo%
\ \BBA {} Scholer%
}{%
{\protect \APACyear {2006}}%
}]{%
matsukiyo_microinstabilities_2006}
\APACinsertmetastar {%
matsukiyo_microinstabilities_2006}%
\begin{APACrefauthors}%
Matsukiyo, S.%
\BCBT {}\ \BBA {} Scholer, M.%
\end{APACrefauthors}%
\unskip\
\newblock
\APACrefYearMonthDay{2006}{}{}.
\newblock
{\BBOQ}\APACrefatitle {On Microinstabilities in the Foot of High {{Mach}}
  Number Perpendicular Shocks} {On microinstabilities in the foot of high
  {{Mach}} number perpendicular shocks}.{\BBCQ}
\newblock
\APACjournalVolNumPages{J. Geophys. Res. Space Phys.}{111}{A6}{}.
\newblock
\begin{APACrefDOI} \doi{10.1029/2005JA011409} \end{APACrefDOI}
\PrintBackRefs{\CurrentBib}

\bibitem [\protect \citeauthoryear {%
Matsumoto%
\ \protect \BOthers {.}}{%
Matsumoto%
\ \protect \BOthers {.}}{%
{\protect \APACyear {1994}}%
}]{%
matsumoto_electrostatic_1994}
\APACinsertmetastar {%
matsumoto_electrostatic_1994}%
\begin{APACrefauthors}%
Matsumoto, H.%
, Kojima, H.%
, Miyatake, T.%
, Omura, Y.%
, Okada, M.%
, Nagano, I.%
\BCBL {}\ \BBA {} Tsutsui, M.%
\end{APACrefauthors}%
\unskip\
\newblock
\APACrefYearMonthDay{1994}{}{}.
\newblock
{\BBOQ}\APACrefatitle {Electrostatic Solitary Waves ({{ESW}}) in the
  Magnetotail: {{BEN}} Wave Forms Observed by {{GEOTAIL}}} {Electrostatic
  solitary waves ({{ESW}}) in the magnetotail: {{BEN}} wave forms observed by
  {{GEOTAIL}}}.{\BBCQ}
\newblock
\APACjournalVolNumPages{Geophys. Res. Lett.}{21}{25}{2915--2918}.
\newblock
\begin{APACrefDOI} \doi{10.1029/94GL01284} \end{APACrefDOI}
\PrintBackRefs{\CurrentBib}

\bibitem [\protect \citeauthoryear {%
McBride%
, Ott%
, Boris%
\BCBL {}\ \BBA {} Orens%
}{%
McBride%
\ \protect \BOthers {.}}{%
{\protect \APACyear {1972}}%
}]{%
mcbride_theory_1972}
\APACinsertmetastar {%
mcbride_theory_1972}%
\begin{APACrefauthors}%
McBride, J\BPBI B.%
, Ott, E.%
, Boris, J\BPBI P.%
\BCBL {}\ \BBA {} Orens, J\BPBI H.%
\end{APACrefauthors}%
\unskip\
\newblock
\APACrefYearMonthDay{1972}{{\APACmonth{12}}}{}.
\newblock
{\BBOQ}\APACrefatitle {Theory and {{Simulation}} of {{Turbulent Heating}} by
  the {{Modified Two}}-{{Stream Instability}}} {Theory and {{Simulation}} of
  {{Turbulent Heating}} by the {{Modified Two}}-{{Stream Instability}}}.{\BBCQ}
\newblock
\APACjournalVolNumPages{The Physics of Fluids}{15}{12}{2367--2383}.
\newblock
\begin{APACrefDOI} \doi{10.1063/1.1693881} \end{APACrefDOI}
\PrintBackRefs{\CurrentBib}

\bibitem [\protect \citeauthoryear {%
McFadden%
\ \protect \BOthers {.}}{%
McFadden%
\ \protect \BOthers {.}}{%
{\protect \APACyear {2008}}%
}]{%
mcfadden_themis_2008}
\APACinsertmetastar {%
mcfadden_themis_2008}%
\begin{APACrefauthors}%
McFadden, J\BPBI P.%
, Carlson, C\BPBI W.%
, Larson, D.%
, Ludlam, M.%
, Abiad, R.%
, Elliott, B.%
\BDBL {}Angelopoulos, V.%
\end{APACrefauthors}%
\unskip\
\newblock
\APACrefYearMonthDay{2008}{{\APACmonth{12}}}{}.
\newblock
{\BBOQ}\APACrefatitle {The {{THEMIS ESA Plasma Instrument}} and {{In}}-Flight
  {{Calibration}}} {The {{THEMIS ESA Plasma Instrument}} and {{In}}-flight
  {{Calibration}}}.{\BBCQ}
\newblock
\APACjournalVolNumPages{Space Sci Rev}{141}{1}{277--302}.
\newblock
\begin{APACrefDOI} \doi{10.1007/s11214-008-9440-2} \end{APACrefDOI}
\PrintBackRefs{\CurrentBib}

\bibitem [\protect \citeauthoryear {%
Mozer%
\ \protect \BOthers {.}}{%
Mozer%
\ \protect \BOthers {.}}{%
{\protect \APACyear {2016}}%
}]{%
mozer_magnetospheric_2016}
\APACinsertmetastar {%
mozer_magnetospheric_2016}%
\begin{APACrefauthors}%
Mozer, F\BPBI S.%
, Agapitov, O\BPBI A.%
, Artemyev, A.%
, Burch, J\BPBI L.%
, Ergun, R\BPBI E.%
, Giles, B\BPBI L.%
\BDBL {}Vasko, I.%
\end{APACrefauthors}%
\unskip\
\newblock
\APACrefYearMonthDay{2016}{{\APACmonth{04}}}{}.
\newblock
{\BBOQ}\APACrefatitle {Magnetospheric {{Multiscale Satellite Observations}} of
  {{Parallel Electron Acceleration}} in {{Magnetic Field Reconnection}} by
  {{Fermi Reflection}} from {{Time Domain Structures}}} {Magnetospheric
  {{Multiscale Satellite Observations}} of {{Parallel Electron Acceleration}}
  in {{Magnetic Field Reconnection}} by {{Fermi Reflection}} from {{Time Domain
  Structures}}}.{\BBCQ}
\newblock
\APACjournalVolNumPages{Phys. Rev. Lett.}{116}{14}{145101}.
\newblock
\begin{APACrefDOI} \doi{10.1103/PhysRevLett.116.145101} \end{APACrefDOI}
\PrintBackRefs{\CurrentBib}

\bibitem [\protect \citeauthoryear {%
Mozer%
\ \protect \BOthers {.}}{%
Mozer%
\ \protect \BOthers {.}}{%
{\protect \APACyear {2015}}%
}]{%
mozer_time_2015}
\APACinsertmetastar {%
mozer_time_2015}%
\begin{APACrefauthors}%
Mozer, F\BPBI S.%
, Agapitov, O\BPBI V.%
, Artemyev, A.%
, Drake, J\BPBI F.%
, Krasnoselskikh, V.%
, Lejosne, S.%
\BCBL {}\ \BBA {} Vasko, I.%
\end{APACrefauthors}%
\unskip\
\newblock
\APACrefYearMonthDay{2015}{}{}.
\newblock
{\BBOQ}\APACrefatitle {Time Domain Structures: {{What}} and Where They Are,
  What They Do, and How They Are Made} {Time domain structures: {{What}} and
  where they are, what they do, and how they are made}.{\BBCQ}
\newblock
\APACjournalVolNumPages{Geophys. Res. Lett.}{42}{10}{3627--3638}.
\newblock
\begin{APACrefDOI} \doi{10.1002/2015GL063946} \end{APACrefDOI}
\PrintBackRefs{\CurrentBib}

\bibitem [\protect \citeauthoryear {%
Muschietti%
\ \BBA {} Lemb{\`e}ge%
}{%
Muschietti%
\ \BBA {} Lemb{\`e}ge%
}{%
{\protect \APACyear {2013}}%
}]{%
muschietti_microturbulence_2013}
\APACinsertmetastar {%
muschietti_microturbulence_2013}%
\begin{APACrefauthors}%
Muschietti, L.%
\BCBT {}\ \BBA {} Lemb{\`e}ge, B.%
\end{APACrefauthors}%
\unskip\
\newblock
\APACrefYearMonthDay{2013}{}{}.
\newblock
{\BBOQ}\APACrefatitle {Microturbulence in the Electron Cyclotron Frequency
  Range at Perpendicular Supercritical Shocks} {Microturbulence in the electron
  cyclotron frequency range at perpendicular supercritical shocks}.{\BBCQ}
\newblock
\APACjournalVolNumPages{J. Geophys. Res. Space Phys.}{118}{5}{2267--2285}.
\newblock
\begin{APACrefDOI} \doi{10.1002/jgra.50224} \end{APACrefDOI}
\PrintBackRefs{\CurrentBib}

\bibitem [\protect \citeauthoryear {%
Nakagawa%
}{%
Nakagawa%
}{%
{\protect \APACyear {2016}}%
}]{%
nakagawa_ulfelf_2016}
\APACinsertmetastar {%
nakagawa_ulfelf_2016}%
\begin{APACrefauthors}%
Nakagawa, T.%
\end{APACrefauthors}%
\unskip\
\newblock
\APACrefYearMonthDay{2016}{}{}.
\newblock
{\BBOQ}\APACrefatitle {{{ULF}}/{{ELF Waves}} in {{Near}}-{{Moon Space}}}
  {{{ULF}}/{{ELF Waves}} in {{Near}}-{{Moon Space}}}.{\BBCQ}
\newblock
\BIn{} \APACrefbtitle {Low-{{Frequency Waves}} in {{Space Plasmas}}}
  {Low-{{Frequency Waves}} in {{Space Plasmas}}}\ (\BPGS\ 293--306).
\newblock
\APACaddressPublisher{}{{American Geophysical Union (AGU)}}.
\newblock
\begin{APACrefDOI} \doi{10.1002/9781119055006.ch17} \end{APACrefDOI}
\PrintBackRefs{\CurrentBib}

\bibitem [\protect \citeauthoryear {%
Reasoner%
\ \BBA {} Burke%
}{%
Reasoner%
\ \BBA {} Burke%
}{%
{\protect \APACyear {1972}}%
}]{%
reasoner_characteristics_1972}
\APACinsertmetastar {%
reasoner_characteristics_1972}%
\begin{APACrefauthors}%
Reasoner, D\BPBI L.%
\BCBT {}\ \BBA {} Burke, W\BPBI J.%
\end{APACrefauthors}%
\unskip\
\newblock
\APACrefYearMonthDay{1972}{}{}.
\newblock
{\BBOQ}\APACrefatitle {Characteristics of the Lunar Photoelectron Layer in the
  Geomagnetic Tail} {Characteristics of the lunar photoelectron layer in the
  geomagnetic tail}.{\BBCQ}
\newblock
\APACjournalVolNumPages{J. Geophys. Res. 1896-1977}{77}{34}{6671--6687}.
\newblock
\begin{APACrefDOI} \doi{10.1029/JA077i034p06671} \end{APACrefDOI}
\PrintBackRefs{\CurrentBib}

\bibitem [\protect \citeauthoryear {%
Ripin%
\ \BBA {} Stenzel%
}{%
Ripin%
\ \BBA {} Stenzel%
}{%
{\protect \APACyear {1973}}%
}]{%
ripin_electron_1973}
\APACinsertmetastar {%
ripin_electron_1973}%
\begin{APACrefauthors}%
Ripin, B\BPBI H.%
\BCBT {}\ \BBA {} Stenzel, R\BPBI L.%
\end{APACrefauthors}%
\unskip\
\newblock
\APACrefYearMonthDay{1973}{{\APACmonth{01}}}{}.
\newblock
{\BBOQ}\APACrefatitle {Electron {{Cyclotron Drift Instability Experiment}}}
  {Electron {{Cyclotron Drift Instability Experiment}}}.{\BBCQ}
\newblock
\APACjournalVolNumPages{Phys. Rev. Lett.}{30}{2}{45--48}.
\newblock
\begin{APACrefDOI} \doi{10.1103/PhysRevLett.30.45} \end{APACrefDOI}
\PrintBackRefs{\CurrentBib}

\bibitem [\protect \citeauthoryear {%
Roux%
\ \protect \BOthers {.}}{%
Roux%
\ \protect \BOthers {.}}{%
{\protect \APACyear {2008}}%
}]{%
roux_search_2008}
\APACinsertmetastar {%
roux_search_2008}%
\begin{APACrefauthors}%
Roux, A.%
, Le~Contel, O.%
, Coillot, C.%
, Bouabdellah, A.%
, {de la Porte}, B.%
, Alison, D.%
\BDBL {}Vassal, M\BPBI C.%
\end{APACrefauthors}%
\unskip\
\newblock
\APACrefYearMonthDay{2008}{{\APACmonth{12}}}{}.
\newblock
{\BBOQ}\APACrefatitle {The {{Search Coil Magnetometer}} for {{THEMIS}}} {The
  {{Search Coil Magnetometer}} for {{THEMIS}}}.{\BBCQ}
\newblock
\APACjournalVolNumPages{Space Sci Rev}{141}{1}{265--275}.
\newblock
\begin{APACrefDOI} \doi{10.1007/s11214-008-9455-8} \end{APACrefDOI}
\PrintBackRefs{\CurrentBib}

\bibitem [\protect \citeauthoryear {%
Russell%
\ \BBA {} Lichtenstein%
}{%
Russell%
\ \BBA {} Lichtenstein%
}{%
{\protect \APACyear {1975}}%
}]{%
russell_source_1975}
\APACinsertmetastar {%
russell_source_1975}%
\begin{APACrefauthors}%
Russell, C\BPBI T.%
\BCBT {}\ \BBA {} Lichtenstein, B\BPBI R.%
\end{APACrefauthors}%
\unskip\
\newblock
\APACrefYearMonthDay{1975}{}{}.
\newblock
{\BBOQ}\APACrefatitle {On the Source of Lunar Limb Compressions} {On the source
  of lunar limb compressions}.{\BBCQ}
\newblock
\APACjournalVolNumPages{J. Geophys. Res. 1896-1977}{80}{34}{4700--4711}.
\newblock
\begin{APACrefDOI} \doi{10.1029/JA080i034p04700} \end{APACrefDOI}
\PrintBackRefs{\CurrentBib}

\bibitem [\protect \citeauthoryear {%
Saito%
\ \protect \BOthers {.}}{%
Saito%
\ \protect \BOthers {.}}{%
{\protect \APACyear {2012}}%
}]{%
saito_simultaneous_2012}
\APACinsertmetastar {%
saito_simultaneous_2012}%
\begin{APACrefauthors}%
Saito, Y.%
, Nishino, M\BPBI N.%
, Fujimoto, M.%
, Yamamoto, T.%
, Yokota, S.%
, Tsunakawa, H.%
\BDBL {}Takahashi, F.%
\end{APACrefauthors}%
\unskip\
\newblock
\APACrefYearMonthDay{2012}{{\APACmonth{03}}}{}.
\newblock
{\BBOQ}\APACrefatitle {Simultaneous Observation of the Electron Acceleration
  and Ion Deceleration over Lunar Magnetic Anomalies} {Simultaneous observation
  of the electron acceleration and ion deceleration over lunar magnetic
  anomalies}.{\BBCQ}
\newblock
\APACjournalVolNumPages{Earth Planet Sp}{64}{2}{4}.
\newblock
\begin{APACrefDOI} \doi{10.5047/eps.2011.07.011} \end{APACrefDOI}
\PrintBackRefs{\CurrentBib}

\bibitem [\protect \citeauthoryear {%
Saito%
\ \protect \BOthers {.}}{%
Saito%
\ \protect \BOthers {.}}{%
{\protect \APACyear {2010}}%
}]{%
saito_-flight_2010}
\APACinsertmetastar {%
saito_-flight_2010}%
\begin{APACrefauthors}%
Saito, Y.%
, Yokota, S.%
, Asamura, K.%
, Tanaka, T.%
, Nishino, M\BPBI N.%
, Yamamoto, T.%
\BDBL {}Takahashi, F.%
\end{APACrefauthors}%
\unskip\
\newblock
\APACrefYearMonthDay{2010}{{\APACmonth{07}}}{}.
\newblock
{\BBOQ}\APACrefatitle {In-Flight {{Performance}} and {{Initial Results}} of
  {{Plasma Energy Angle}} and {{Composition Experiment}} ({{PACE}})
  on~{{SELENE}} ({{Kaguya}})} {In-flight {{Performance}} and {{Initial
  Results}} of {{Plasma Energy Angle}} and {{Composition Experiment}}
  ({{PACE}}) on~{{SELENE}} ({{Kaguya}})}.{\BBCQ}
\newblock
\APACjournalVolNumPages{Space Sci Rev}{154}{1}{265--303}.
\newblock
\begin{APACrefDOI} \doi{10.1007/s11214-010-9647-x} \end{APACrefDOI}
\PrintBackRefs{\CurrentBib}

\bibitem [\protect \citeauthoryear {%
Steinvall%
\ \protect \BOthers {.}}{%
Steinvall%
\ \protect \BOthers {.}}{%
{\protect \APACyear {2019}}%
}]{%
steinvall_multispacecraft_2019}
\APACinsertmetastar {%
steinvall_multispacecraft_2019}%
\begin{APACrefauthors}%
Steinvall, K.%
, Khotyaintsev, Y\BPBI V.%
, Graham, D\BPBI B.%
, Vaivads, A.%
, Lindqvist, P\BHBI A.%
, Russell, C\BPBI T.%
\BCBL {}\ \BBA {} Burch, J\BPBI L.%
\end{APACrefauthors}%
\unskip\
\newblock
\APACrefYearMonthDay{2019}{}{}.
\newblock
{\BBOQ}\APACrefatitle {Multispacecraft {{Analysis}} of {{Electron Holes}}}
  {Multispacecraft {{Analysis}} of {{Electron Holes}}}.{\BBCQ}
\newblock
\APACjournalVolNumPages{Geophys. Res. Lett.}{46}{1}{55--63}.
\newblock
\begin{APACrefDOI} \doi{10.1029/2018GL080757} \end{APACrefDOI}
\PrintBackRefs{\CurrentBib}

\bibitem [\protect \citeauthoryear {%
Stenzel%
}{%
Stenzel%
}{%
{\protect \APACyear {2016}}%
}]{%
stenzel_whistler_2016}
\APACinsertmetastar {%
stenzel_whistler_2016}%
\begin{APACrefauthors}%
Stenzel, R\BPBI L.%
\end{APACrefauthors}%
\unskip\
\newblock
\APACrefYearMonthDay{2016}{{\APACmonth{07}}}{}.
\newblock
{\BBOQ}\APACrefatitle {Whistler Waves with Angular Momentum in Space and
  Laboratory Plasmas and Their Counterparts in Free Space} {Whistler waves with
  angular momentum in space and laboratory plasmas and their counterparts in
  free space}.{\BBCQ}
\newblock
\APACjournalVolNumPages{Adv. Phys. X}{1}{4}{687--710}.
\newblock
\begin{APACrefDOI} \doi{10.1080/23746149.2016.1240017} \end{APACrefDOI}
\PrintBackRefs{\CurrentBib}

\bibitem [\protect \citeauthoryear {%
Stenzel%
\ \BBA {} Ripin%
}{%
Stenzel%
\ \BBA {} Ripin%
}{%
{\protect \APACyear {1973}}%
}]{%
stenzel_growth_1973}
\APACinsertmetastar {%
stenzel_growth_1973}%
\begin{APACrefauthors}%
Stenzel, R\BPBI L.%
\BCBT {}\ \BBA {} Ripin, B\BPBI H.%
\end{APACrefauthors}%
\unskip\
\newblock
\APACrefYearMonthDay{1973}{{\APACmonth{12}}}{}.
\newblock
{\BBOQ}\APACrefatitle {Growth and {{Saturation}} of the {{Absolute
  Electron}}-{{Cyclotron Drift Instability}}} {Growth and {{Saturation}} of the
  {{Absolute Electron}}-{{Cyclotron Drift Instability}}}.{\BBCQ}
\newblock
\APACjournalVolNumPages{Phys. Rev. Lett.}{31}{26}{1545--1548}.
\newblock
\begin{APACrefDOI} \doi{10.1103/PhysRevLett.31.1545} \end{APACrefDOI}
\PrintBackRefs{\CurrentBib}

\bibitem [\protect \citeauthoryear {%
Tsugawa%
\ \protect \BOthers {.}}{%
Tsugawa%
\ \protect \BOthers {.}}{%
{\protect \APACyear {2011}}%
}]{%
tsugawa_statistical_2011}
\APACinsertmetastar {%
tsugawa_statistical_2011}%
\begin{APACrefauthors}%
Tsugawa, Y.%
, Terada, N.%
, Katoh, Y.%
, Ono, T.%
, Tsunakawa, H.%
, Takahashi, F.%
\BDBL {}Matsushima, M.%
\end{APACrefauthors}%
\unskip\
\newblock
\APACrefYearMonthDay{2011}{{\APACmonth{05}}}{}.
\newblock
{\BBOQ}\APACrefatitle {Statistical Analysis of Monochromatic Whistler Waves
  near the {{Moon}} Detected by {{Kaguya}}} {Statistical analysis of
  monochromatic whistler waves near the {{Moon}} detected by
  {{Kaguya}}}.{\BBCQ}
\newblock
\APACjournalVolNumPages{Ann. Geophys.}{29}{5}{889--893}.
\newblock
\begin{APACrefDOI} \doi{10.5194/angeo-29-889-2011} \end{APACrefDOI}
\PrintBackRefs{\CurrentBib}

\bibitem [\protect \citeauthoryear {%
Tsunakawa%
, Takahashi%
, Shimizu%
, Shibuya%
\BCBL {}\ \BBA {} Matsushima%
}{%
Tsunakawa%
\ \protect \BOthers {.}}{%
{\protect \APACyear {2015}}%
}]{%
tsunakawa_surface_2015}
\APACinsertmetastar {%
tsunakawa_surface_2015}%
\begin{APACrefauthors}%
Tsunakawa, H.%
, Takahashi, F.%
, Shimizu, H.%
, Shibuya, H.%
\BCBL {}\ \BBA {} Matsushima, M.%
\end{APACrefauthors}%
\unskip\
\newblock
\APACrefYearMonthDay{2015}{}{}.
\newblock
{\BBOQ}\APACrefatitle {Surface Vector Mapping of Magnetic Anomalies over the
  {{Moon}} Using {{Kaguya}} and {{Lunar Prospector}} Observations} {Surface
  vector mapping of magnetic anomalies over the {{Moon}} using {{Kaguya}} and
  {{Lunar Prospector}} observations}.{\BBCQ}
\newblock
\APACjournalVolNumPages{J. Geophys. Res. Planets}{120}{6}{1160--1185}.
\newblock
\begin{APACrefDOI} \doi{10.1002/2014JE004785} \end{APACrefDOI}
\PrintBackRefs{\CurrentBib}

\bibitem [\protect \citeauthoryear {%
Tsurutani%
\ \BBA {} Lakhina%
}{%
Tsurutani%
\ \BBA {} Lakhina%
}{%
{\protect \APACyear {1997}}%
}]{%
tsurutani_basic_1997}
\APACinsertmetastar {%
tsurutani_basic_1997}%
\begin{APACrefauthors}%
Tsurutani, B\BPBI T.%
\BCBT {}\ \BBA {} Lakhina, G\BPBI S.%
\end{APACrefauthors}%
\unskip\
\newblock
\APACrefYearMonthDay{1997}{}{}.
\newblock
{\BBOQ}\APACrefatitle {Some Basic Concepts of Wave-Particle Interactions in
  Collisionless Plasmas} {Some basic concepts of wave-particle interactions in
  collisionless plasmas}.{\BBCQ}
\newblock
\APACjournalVolNumPages{Rev. Geophys.}{35}{4}{491--501}.
\newblock
\begin{APACrefDOI} \doi{10.1029/97RG02200} \end{APACrefDOI}
\PrintBackRefs{\CurrentBib}

\bibitem [\protect \citeauthoryear {%
Vasko%
\ \protect \BOthers {.}}{%
Vasko%
\ \protect \BOthers {.}}{%
{\protect \APACyear {2017}}%
}]{%
vasko_diffusive_2017}
\APACinsertmetastar {%
vasko_diffusive_2017}%
\begin{APACrefauthors}%
Vasko, I\BPBI Y.%
, Agapitov, O\BPBI V.%
, Mozer, F\BPBI S.%
, Artemyev, A\BPBI V.%
, Krasnoselskikh, V\BPBI V.%
\BCBL {}\ \BBA {} Bonnell, J\BPBI W.%
\end{APACrefauthors}%
\unskip\
\newblock
\APACrefYearMonthDay{2017}{}{}.
\newblock
{\BBOQ}\APACrefatitle {Diffusive Scattering of Electrons by Electron Holes
  around Injection Fronts} {Diffusive scattering of electrons by electron holes
  around injection fronts}.{\BBCQ}
\newblock
\APACjournalVolNumPages{J. Geophys. Res. Space Phys.}{122}{3}{3163--3182}.
\newblock
\begin{APACrefDOI} \doi{10.1002/2016JA023337} \end{APACrefDOI}
\PrintBackRefs{\CurrentBib}

\bibitem [\protect \citeauthoryear {%
Vorburger%
\ \protect \BOthers {.}}{%
Vorburger%
\ \protect \BOthers {.}}{%
{\protect \APACyear {2012}}%
}]{%
vorburger_energetic_2012}
\APACinsertmetastar {%
vorburger_energetic_2012}%
\begin{APACrefauthors}%
Vorburger, A.%
, Wurz, P.%
, Barabash, S.%
, Wieser, M.%
, Futaana, Y.%
, Holmstr{\"o}m, M.%
\BDBL {}Asamura, K.%
\end{APACrefauthors}%
\unskip\
\newblock
\APACrefYearMonthDay{2012}{}{}.
\newblock
{\BBOQ}\APACrefatitle {Energetic Neutral Atom Observations of Magnetic
  Anomalies on the Lunar Surface} {Energetic neutral atom observations of
  magnetic anomalies on the lunar surface}.{\BBCQ}
\newblock
\APACjournalVolNumPages{J. Geophys. Res. Space Phys.}{117}{A7}{}.
\newblock
\begin{APACrefDOI} \doi{10.1029/2012JA017553} \end{APACrefDOI}
\PrintBackRefs{\CurrentBib}

\bibitem [\protect \citeauthoryear {%
Whipple%
}{%
Whipple%
}{%
{\protect \APACyear {1981}}%
}]{%
whipple_potentials_1981}
\APACinsertmetastar {%
whipple_potentials_1981}%
\begin{APACrefauthors}%
Whipple, E\BPBI C.%
\end{APACrefauthors}%
\unskip\
\newblock
\APACrefYearMonthDay{1981}{{\APACmonth{11}}}{}.
\newblock
{\BBOQ}\APACrefatitle {Potentials of Surfaces in Space} {Potentials of surfaces
  in space}.{\BBCQ}
\newblock
\APACjournalVolNumPages{Rep. Prog. Phys.}{44}{11}{1197--1250}.
\newblock
\begin{APACrefDOI} \doi{10.1088/0034-4885/44/11/002} \end{APACrefDOI}
\PrintBackRefs{\CurrentBib}

\bibitem [\protect \citeauthoryear {%
Wieczorek%
}{%
Wieczorek%
}{%
{\protect \APACyear {2018}}%
}]{%
wieczorek_strength_2018}
\APACinsertmetastar {%
wieczorek_strength_2018}%
\begin{APACrefauthors}%
Wieczorek, M\BPBI A.%
\end{APACrefauthors}%
\unskip\
\newblock
\APACrefYearMonthDay{2018}{}{}.
\newblock
{\BBOQ}\APACrefatitle {Strength, {{Depth}}, and {{Geometry}} of {{Magnetic
  Sources}} in the {{Crust}} of the {{Moon From Localized Power Spectrum
  Analysis}}} {Strength, {{Depth}}, and {{Geometry}} of {{Magnetic Sources}} in
  the {{Crust}} of the {{Moon From Localized Power Spectrum Analysis}}}.{\BBCQ}
\newblock
\APACjournalVolNumPages{J. Geophys. Res. Planets}{123}{1}{291--316}.
\newblock
\begin{APACrefDOI} \doi{10.1002/2017JE005418} \end{APACrefDOI}
\PrintBackRefs{\CurrentBib}

\bibitem [\protect \citeauthoryear {%
Wieser%
\ \protect \BOthers {.}}{%
Wieser%
\ \protect \BOthers {.}}{%
{\protect \APACyear {2010}}%
}]{%
wieser_first_2010}
\APACinsertmetastar {%
wieser_first_2010}%
\begin{APACrefauthors}%
Wieser, M.%
, Barabash, S.%
, Futaana, Y.%
, Holmstr{\"o}m, M.%
, Bhardwaj, A.%
, Sridharan, R.%
\BDBL {}Asamura, K.%
\end{APACrefauthors}%
\unskip\
\newblock
\APACrefYearMonthDay{2010}{}{}.
\newblock
{\BBOQ}\APACrefatitle {First Observation of a Mini-Magnetosphere above a Lunar
  Magnetic Anomaly Using Energetic Neutral Atoms} {First observation of a
  mini-magnetosphere above a lunar magnetic anomaly using energetic neutral
  atoms}.{\BBCQ}
\newblock
\APACjournalVolNumPages{Geophys. Res. Lett.}{37}{5}{}.
\newblock
\begin{APACrefDOI} \doi{10.1029/2009GL041721} \end{APACrefDOI}
\PrintBackRefs{\CurrentBib}

\bibitem [\protect \citeauthoryear {%
Wilson%
\ \protect \BOthers {.}}{%
Wilson%
\ \protect \BOthers {.}}{%
{\protect \APACyear {2010}}%
}]{%
wilson_large-amplitude_2010}
\APACinsertmetastar {%
wilson_large-amplitude_2010}%
\begin{APACrefauthors}%
Wilson, L\BPBI B.%
, Cattell, C\BPBI A.%
, Kellogg, P\BPBI J.%
, Goetz, K.%
, Kersten, K.%
, Kasper, J\BPBI C.%
\BDBL {}Wilber, M.%
\end{APACrefauthors}%
\unskip\
\newblock
\APACrefYearMonthDay{2010}{}{}.
\newblock
{\BBOQ}\APACrefatitle {Large-Amplitude Electrostatic Waves Observed at a
  Supercritical Interplanetary Shock} {Large-amplitude electrostatic waves
  observed at a supercritical interplanetary shock}.{\BBCQ}
\newblock
\APACjournalVolNumPages{J. Geophys. Res. Space Phys.}{115}{A12}{}.
\newblock
\begin{APACrefDOI} \doi{10.1029/2010JA015332} \end{APACrefDOI}
\PrintBackRefs{\CurrentBib}

\bibitem [\protect \citeauthoryear {%
Wong%
}{%
Wong%
}{%
{\protect \APACyear {1970}}%
}]{%
wong_electrostatic_1970}
\APACinsertmetastar {%
wong_electrostatic_1970}%
\begin{APACrefauthors}%
Wong, H\BPBI V.%
\end{APACrefauthors}%
\unskip\
\newblock
\APACrefYearMonthDay{1970}{{\APACmonth{03}}}{}.
\newblock
{\BBOQ}\APACrefatitle {Electrostatic {{Electron}}-{{Ion Streaming
  Instability}}} {Electrostatic {{Electron}}-{{Ion Streaming
  Instability}}}.{\BBCQ}
\newblock
\APACjournalVolNumPages{The Physics of Fluids}{13}{3}{757--760}.
\newblock
\begin{APACrefDOI} \doi{10.1063/1.1692983} \end{APACrefDOI}
\PrintBackRefs{\CurrentBib}

\bibitem [\protect \citeauthoryear {%
Wu%
\ \BBA {} Fredricks%
}{%
Wu%
\ \BBA {} Fredricks%
}{%
{\protect \APACyear {1972}}%
}]{%
wu_cyclotron_1972}
\APACinsertmetastar {%
wu_cyclotron_1972}%
\begin{APACrefauthors}%
Wu, C\BPBI S.%
\BCBT {}\ \BBA {} Fredricks, R\BPBI W.%
\end{APACrefauthors}%
\unskip\
\newblock
\APACrefYearMonthDay{1972}{}{}.
\newblock
{\BBOQ}\APACrefatitle {Cyclotron Drift Instability in the Bow Shock} {Cyclotron
  drift instability in the bow shock}.{\BBCQ}
\newblock
\APACjournalVolNumPages{J. Geophys. Res. 1896-1977}{77}{28}{5585--5589}.
\newblock
\begin{APACrefDOI} \doi{10.1029/JA077i028p05585} \end{APACrefDOI}
\PrintBackRefs{\CurrentBib}

\bibitem [\protect \citeauthoryear {%
Wu%
\ \protect \BOthers {.}}{%
Wu%
\ \protect \BOthers {.}}{%
{\protect \APACyear {1984}}%
}]{%
wu_microinstabilities_1984}
\APACinsertmetastar {%
wu_microinstabilities_1984}%
\begin{APACrefauthors}%
Wu, C\BPBI S.%
, Winske, D.%
, Zhou, Y\BPBI M.%
, Tsai, S\BPBI T.%
, Rodriguez, P.%
, Tanaka, M.%
\BDBL {}Goodrich, C\BPBI C.%
\end{APACrefauthors}%
\unskip\
\newblock
\APACrefYearMonthDay{1984}{{\APACmonth{01}}}{}.
\newblock
{\BBOQ}\APACrefatitle {Microinstabilities Associated with a High {{Mach}}
  Number, Perpendicular Bow Shock} {Microinstabilities associated with a high
  {{Mach}} number, perpendicular bow shock}.{\BBCQ}
\newblock
\APACjournalVolNumPages{Space Sci Rev}{37}{1}{63--109}.
\newblock
\begin{APACrefDOI} \doi{10.1007/BF00213958} \end{APACrefDOI}
\PrintBackRefs{\CurrentBib}

\bibitem [\protect \citeauthoryear {%
Yokota%
\ \protect \BOthers {.}}{%
Yokota%
\ \protect \BOthers {.}}{%
{\protect \APACyear {2009}}%
}]{%
yokota_first_2009}
\APACinsertmetastar {%
yokota_first_2009}%
\begin{APACrefauthors}%
Yokota, S.%
, Saito, Y.%
, Asamura, K.%
, Tanaka, T.%
, Nishino, M\BPBI N.%
, Tsunakawa, H.%
\BDBL {}Terasawa, T.%
\end{APACrefauthors}%
\unskip\
\newblock
\APACrefYearMonthDay{2009}{}{}.
\newblock
{\BBOQ}\APACrefatitle {First Direct Detection of Ions Originating from the
  {{Moon}} by {{MAP}}-{{PACE IMA}} Onboard {{SELENE}} ({{KAGUYA}})} {First
  direct detection of ions originating from the {{Moon}} by {{MAP}}-{{PACE
  IMA}} onboard {{SELENE}} ({{KAGUYA}})}.{\BBCQ}
\newblock
\APACjournalVolNumPages{Geophys. Res. Lett.}{36}{11}{}.
\newblock
\begin{APACrefDOI} \doi{10.1029/2009GL038185} \end{APACrefDOI}
\PrintBackRefs{\CurrentBib}

\end{thebibliography}

%
%
%
%
%

\end{document}